\documentclass{aastex}

\def\ges{\;_\sim^>\;}
\def\les{\;_\sim^<\;}

\slugcomment{Accepted to Astrophysical Journal}

\shorttitle{Guerra, Daly, \& Wan}
\shortauthors{Cosmological Parameters from Radio Galaxies}

\begin{document}

\title{Global Cosmological Parameters Determined 
Using Classical Double Radio Galaxies}

\author{Erick J. Guerra}
\affil{Department of Chemistry \& Physics, Rowan University}
\affil{201 Mullica Hill Road, Glassboro, NJ 08028}
\email{guerra@scherzo.rowan.edu}
\author{Ruth A. Daly}
\affil{Department of Physics, Pennsylvania State University}
\affil{PO Box 7009, Reading, PA 19610-6009}
\email{rdaly@psu.edu}
\and 
\author{Lin Wan}
\affil{Axolotl Corporation}
\affil{800 El Camino Real West, Mountain View, CA 94041}

\begin{abstract}
A sample of 20 powerful extended radio galaxies with redshifts
between zero and two were used to determine constraints on 
global cosmological parameters.  Data for
six radio sources were obtained from the VLA archive,
analyzed, 
and combined with the sample of 14 radio galaxies used 
previously by Guerra \& Daly to determine cosmological
parameters.  The new results are consistent with our previous
results, and indicate that the current value of the 
mean mass density of the  universe
is significantly less than the critical value.  A universe with 
$\Omega_m$ in matter of unity is ruled out at 99.0\% 
confidence, and the best fitting values of $\Omega_m$ in matter
are $0.10^{+0.25}_{-0.10}$ and $-0.25^{+0.35}_{-0.25}$  assuming 
zero space curvature and zero cosmological
constant, respectively.  Note that identical results obtain 
when the low redshift bin, which includes Cygnus A, is excluded; these
results are independent of whether the radio source Cygnus A is included.
The method does not rely on a zero-redshift normalization.

The radio properties of each source are also used to determine
the density of the gas in the vicinity of the source, and
the beam power of the source.  
The six new radio sources have physical characteristics
similar to those found for the original 14 sources.  
The density of the gas around these radio sources
is typical of gas in present day clusters of galaxies.  
The beam powers are typically about $10^{45} \hbox{ erg s}^{-1}$.  

\end{abstract}


\section{Introduction}

The future and ultimate fate of the universe can be predicted 
given a knowledge of the recent expansion history of the
universe (assuming the universe is homogeneous and isotropic on 
scales greater than the current horizon size).  
This recent expansion history can be probed by 
studying the coordinate distance to sources at redshifts of
one or two; the coordinate distance is equivalent to the
luminosity distance or angular size distance multiplied by factors
of $(1+z)$.  The advantage of determining cosmological parameters
using the coordinate distance (luminosity distance or 
angular size distance) is that this distance depends on
global, or average, cosmological parameters.  It is independent
of the way the matter is distributed spatially, of the 
power spectrum of density fluctuations, of whether the matter
is biased relative to the light, and of the form or nature
of the dark matter (assuming that the universe is homogeneous
and isotropic on large scales).  

It was shown in 1994 that powerful double-lobed radio galaxies provide
a modified standard yardstick that can be used to determine
global cosmological parameters (Daly 1994, 1995), much like supernovae
can be used as modified standard candles.  The method was 
applied and discussed in detail by Guerra \& Daly (1996, 1998), Guerra
(1997), and Daly, Guerra, \& Wan (1998) who found that the data 
strongly favor a low density universe; a universe with $\Omega_m
=1$ was ruled out at 97.5 \% confidence.    

It was shown by Daly (1994, 1995) that the radio properties of
these
sources could be used not only to study global cosmological
parameters, but also to determine the ambient gas density, 
beam power, Mach number of lobe advance, and ambient gas
temperature of the sources and their environments.  The 
characteristics of the sources and their environments are 
presented and discussed in a series of papers (Wellman \& Daly
1996a,b; Wan, Daly, \& Wellman 1996; Daly 1996;  
Wellman, Daly,
\& Wan 1997a,b; Wan \& Daly 1998a,b; Wan, Daly, \& Guerra 2000).  

Radio maps of six powerful double-lobed radio galaxies were extracted from the 
Very Large Array (VLA) archives 
at the National Radio Astronomy Observatory (NRAO), and analyzed in detail.  
New results on global cosmological parameters, ambient gas
densities, and beam powers are presented here.  

The expanded sample is described in \S 2.  The new results are
presented in \S 3.   The implications of the results are
discussed in \S 4.  

\section{Expanded Sample}\label{sec:samp}

Each powerful extended radio galaxy (also known as a ``classical double'') 
has a characteristic size, $D_*$, that predicts the lobe-lobe
separation at the end of the source lifetime 
(Daly 1994, 1995; Guerra \& Daly 1998).
The parameters needed to compute the characteristic size
are the lobe
propagation velocity, $v_L$, the lobe width, $a_L$, and the
lobe magnetic field strength, $B_L$.
These three parameters can be determined using 
radio maps with arc-second resolution at multiple frequencies,
such as those produced with the 
VLA or MERLIN (Multi-Element Radio Linked Interferometer Network).  
Multiple-frequency data
are needed to use the theory of spectral aging to estimate
the lobe propagation velocity (e.g., Myers \& Spangler 1985).  In addition, 
these maps
must have the necessary angular resolution and dynamic
range to image sufficient portions of the radio bridges.

Two published
data sets, Leahy, Muxlow, \& Stephens
(1989) and Liu, Pooley, \& Riley (1992),
have radio maps of powerful extended radio galaxies
at multiple frequencies which are sufficient
to compute all three parameters used in determining $D_*$.
These data were used
to compute $D_*$
for 14 radio galaxies
(Guerra \& Daly 1996, 1998; Guerra 1997; Daly, Guerra, \& Wan 1998).
Current efforts to expand the data set are underway, and include
searches through the VLA archive.  The VLA archive search
has already yielded the desired data for six radio galaxies, and new
results including these six sources are presented here.

Data were selected from observations
of powerful extended radio galaxies from the 3CR sample 
(Bennett 1962) on the basis of the observation frequency and
array configuration used.  An observation in the VLA archive was
a candidate if the lobe-lobe angular size of the source was 10 to 40
times the implied beam size, and four hours separated the first and 
last scans.  These observations should
resolve the source sufficiently and have enough $uv$-coverage
to image the bridges.  From candidate observations at
both L and C band, we have 
successfully imaged data from six sources.  The VLA 
archive data sets used here
are listed in Table \ref{tab:arch}.  

Radio imaging was performed using the NRAO
${\cal AIPS}$ software package.
The $uv$ data needed minimal editing, and initial calibration
was performed in the standard manner using ${\cal AIPS}$. 
As diagnostics, initial images were made using the ${\cal AIPS}$
task {\tt IMAGR} both without and with the {\tt CLEAN}
deconvolution algorithm.  The final radio maps 
were produced with the {\tt SCMAP} task, which performs
self-calibration along with the 
{\tt IMAGR} and {\tt CLEAN} tasks.    

Parameters 
used in ${\cal AIPS}$ for imaging, deconvolution, and calibration are
chosen to produce radio maps for a given source that can 
easily be used for spectral aging analysis.
This is particularly important where observations
at different frequencies are produced by different observers.
Although some of these radio maps exist,
different choices of parameters (such as the restoring beam)
are often used
to produce these radio maps.  Thus, reducing
the raw $uv$ data insures that the data analysis can
be performed consistently between radio maps.

\subsection{Radio Maps from Archive Data and Their Use}

\begin{figure}
\epsscale{0.6}
\centerline{(a)}
\plotone{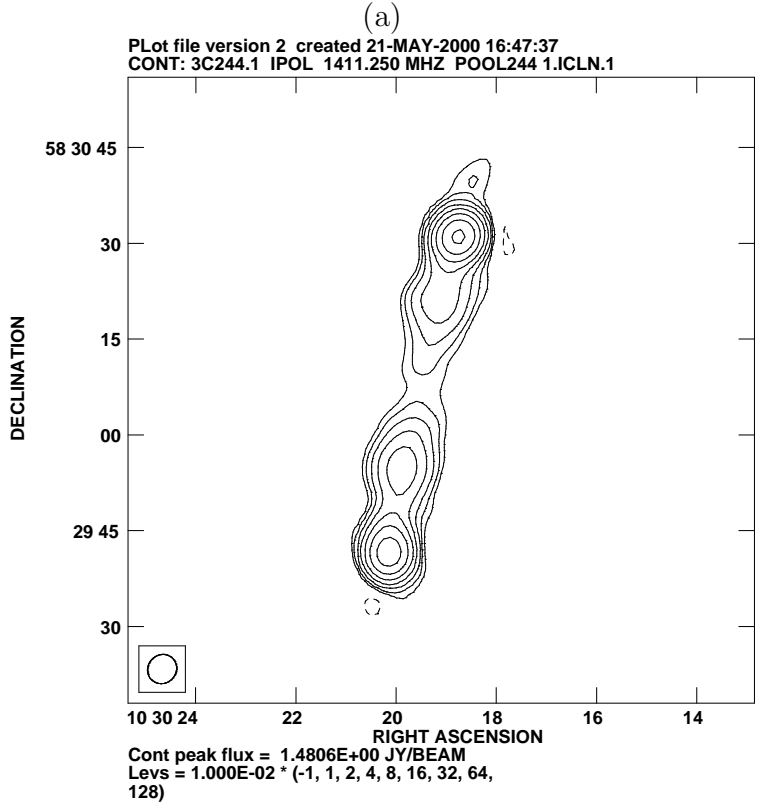}
\centerline{(b)}
\plotone{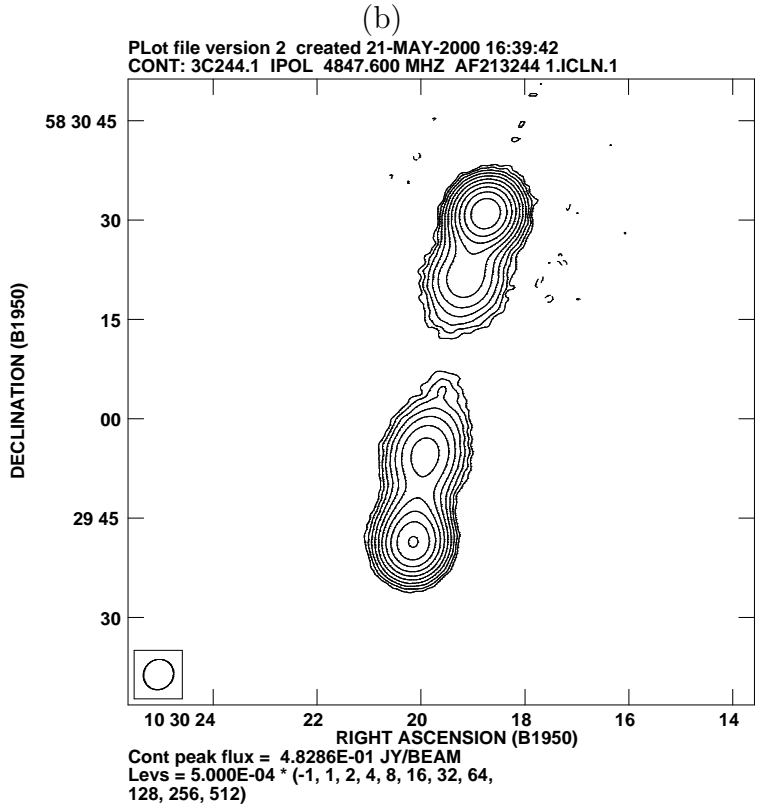}
\caption{Radio maps of 3C 244.1 from archive data 
in (a) L band and (b) C band. \label{fig:2441}}
\end{figure}

\begin{figure}
\epsscale{0.85}
\centerline{(a)}
\plotone{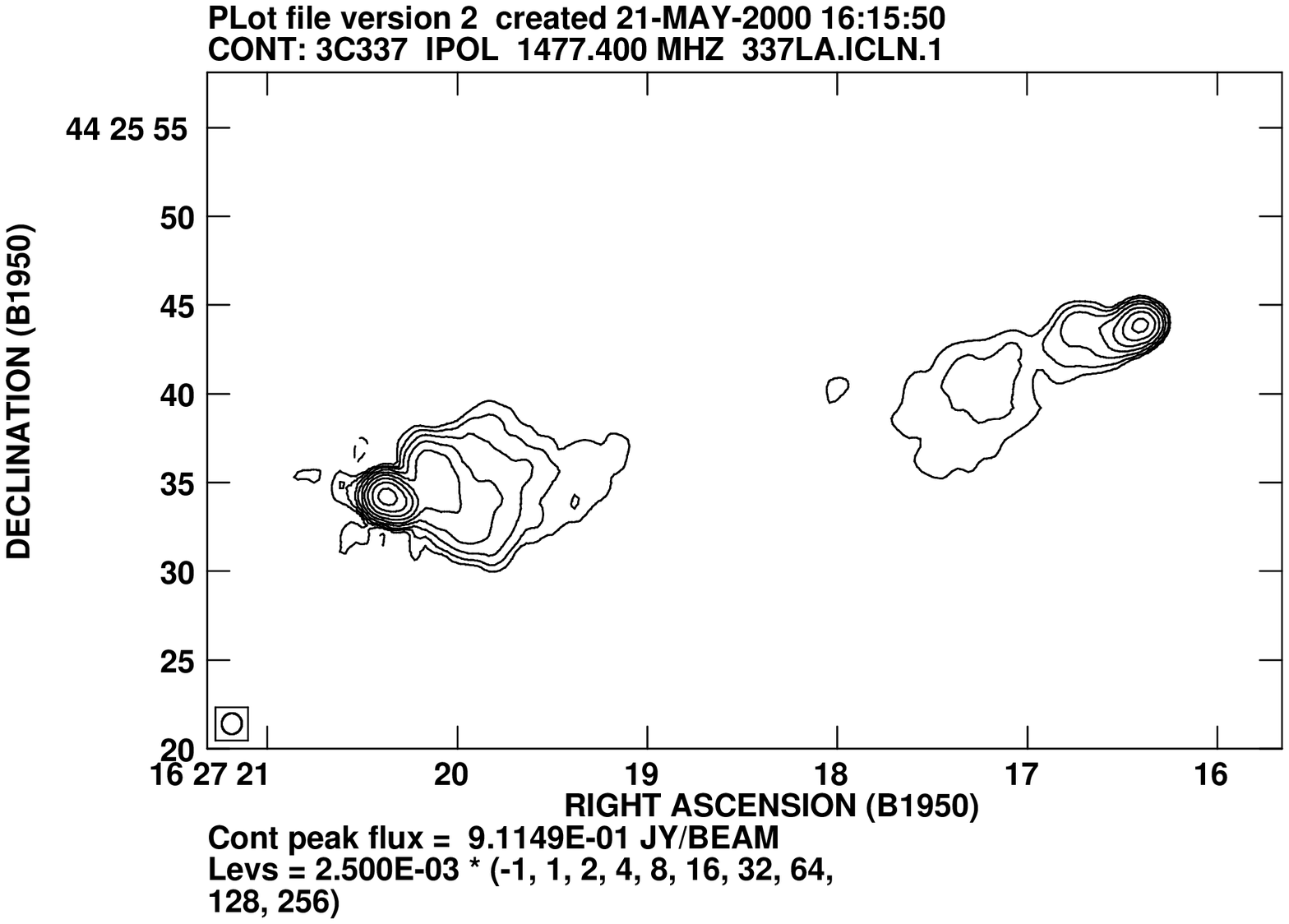}
\centerline{(b)}
\plotone{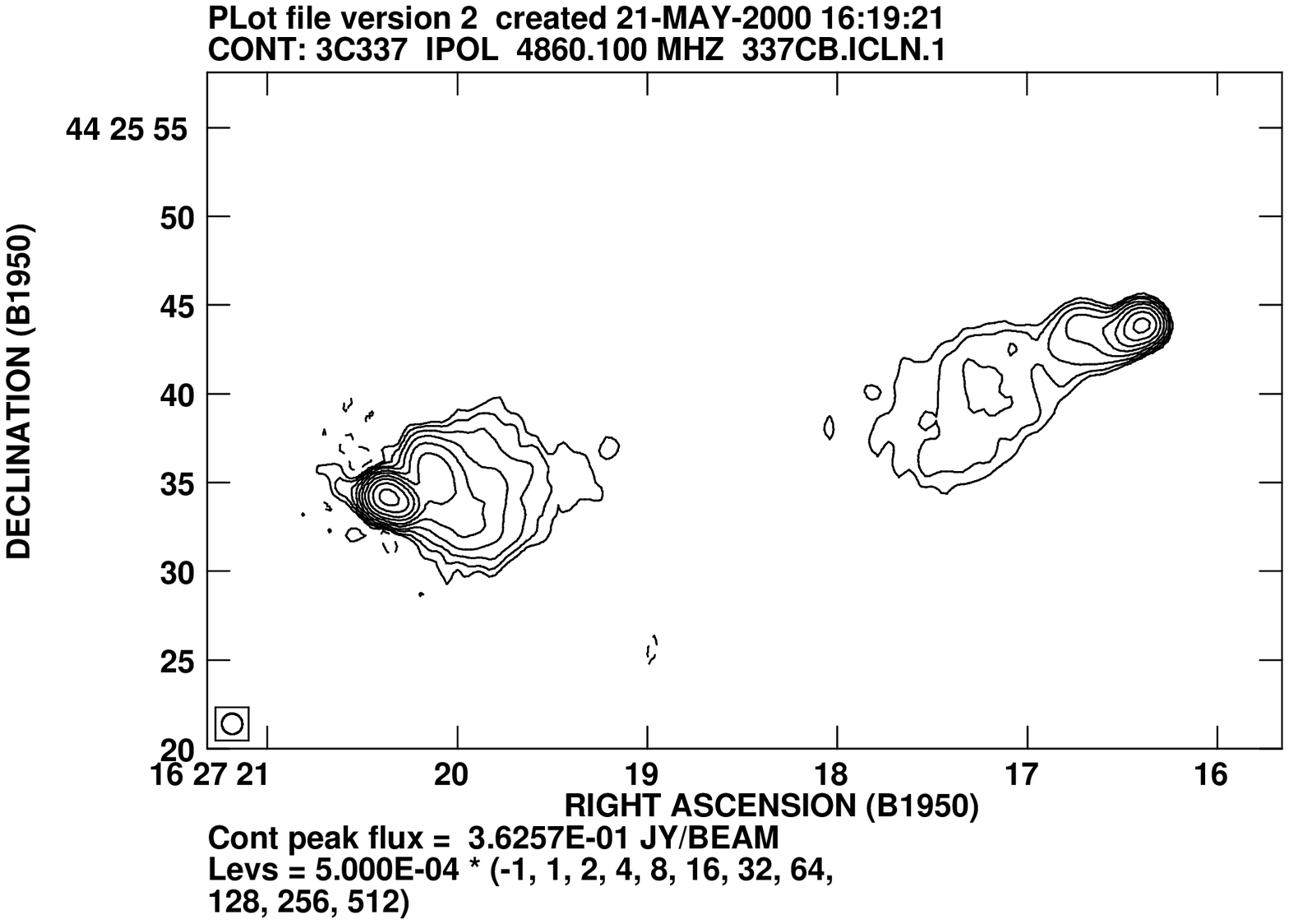}
\caption{Radio maps of 3C 337 from archive data 
in (a) L band and (b) C band. \label{fig:337}}
\end{figure}

\begin{figure}
\epsscale{0.6}
\centerline{(a)}
\plotone{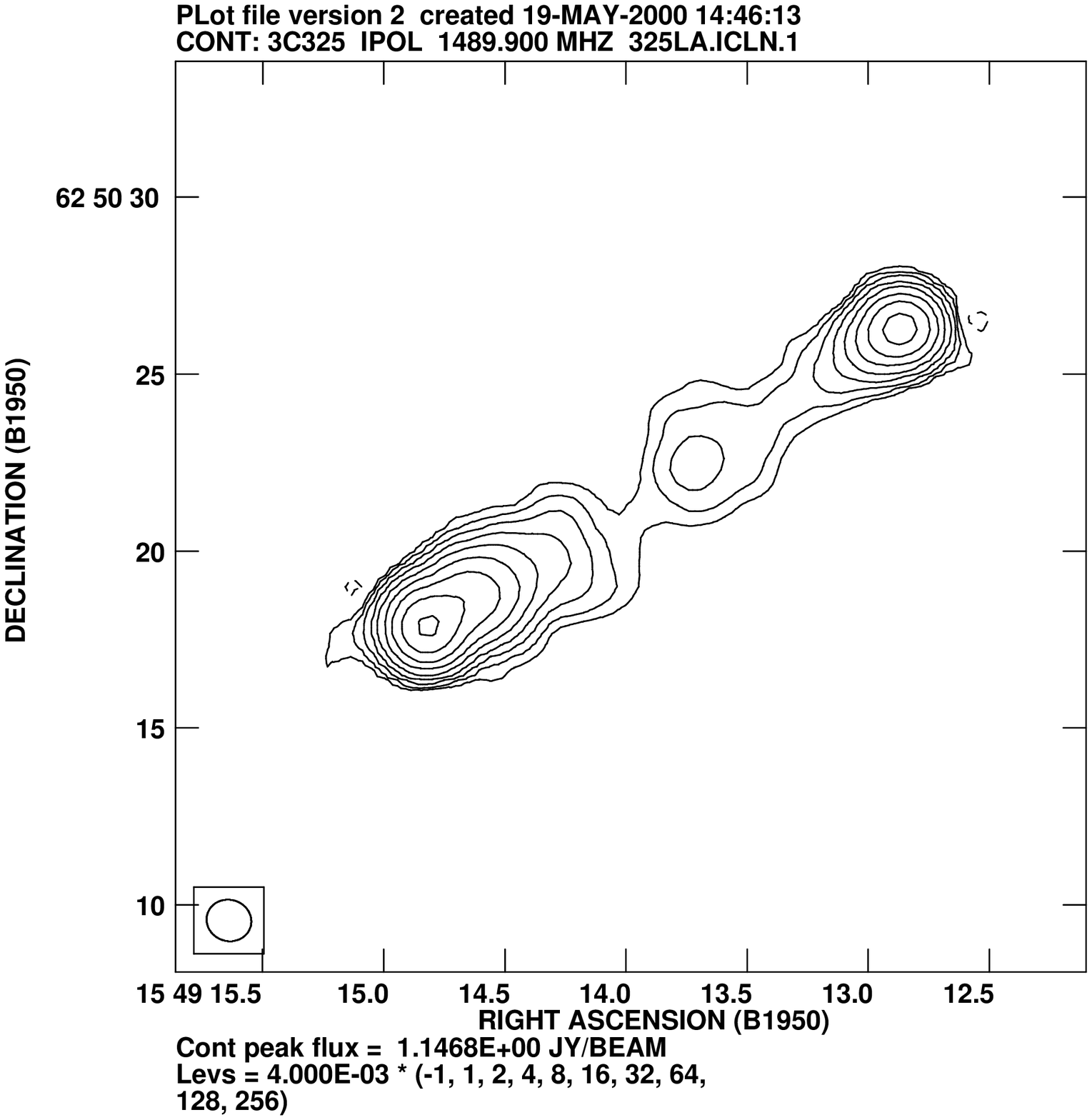}
\centerline{(b)}
\plotone{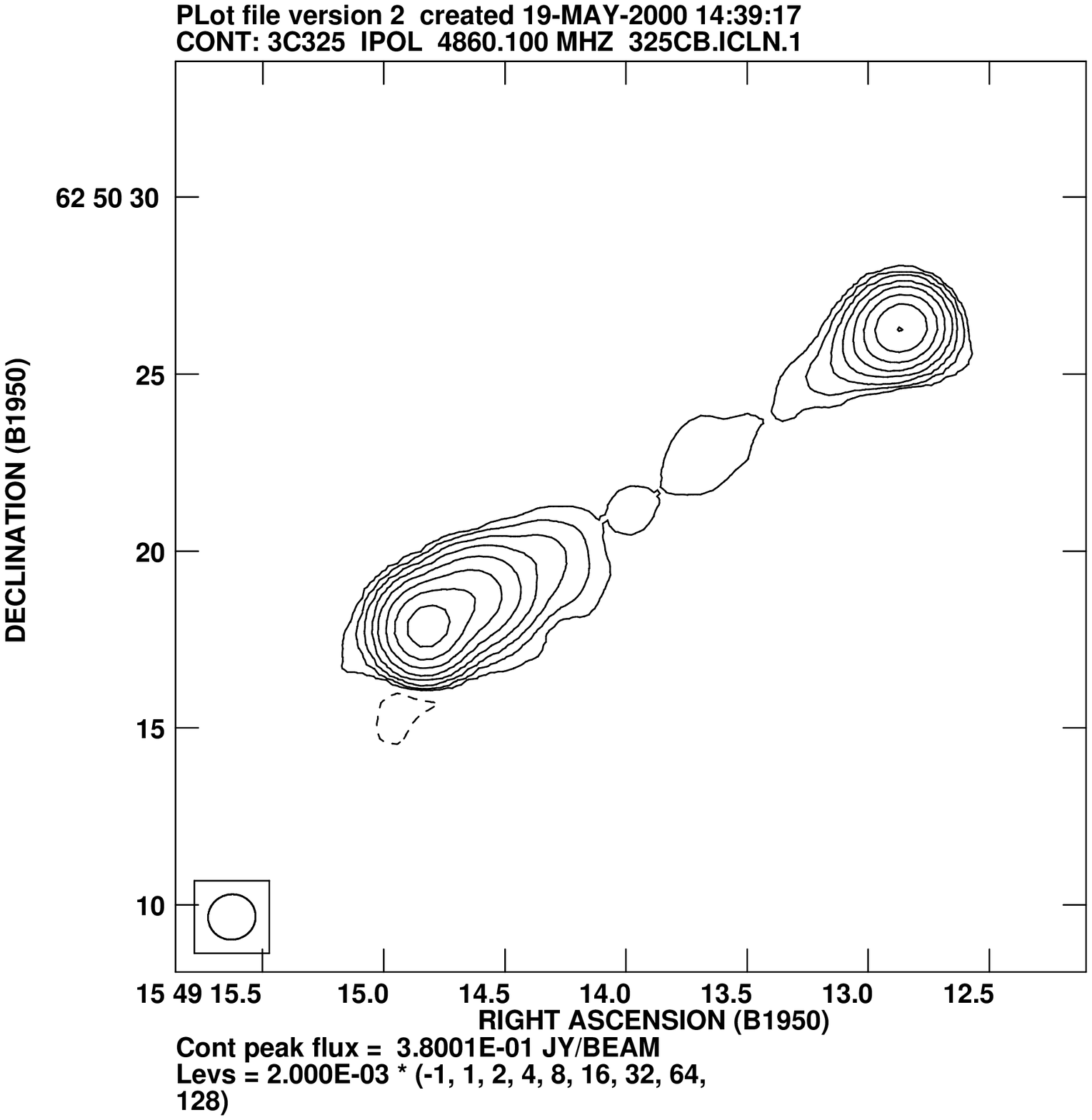}
\caption{Radio maps of 3C 325 from archive data 
in (a) L band and (b) C band. \label{fig:325}}
\end{figure}

\begin{figure}
\centerline{(a)}
\plotone{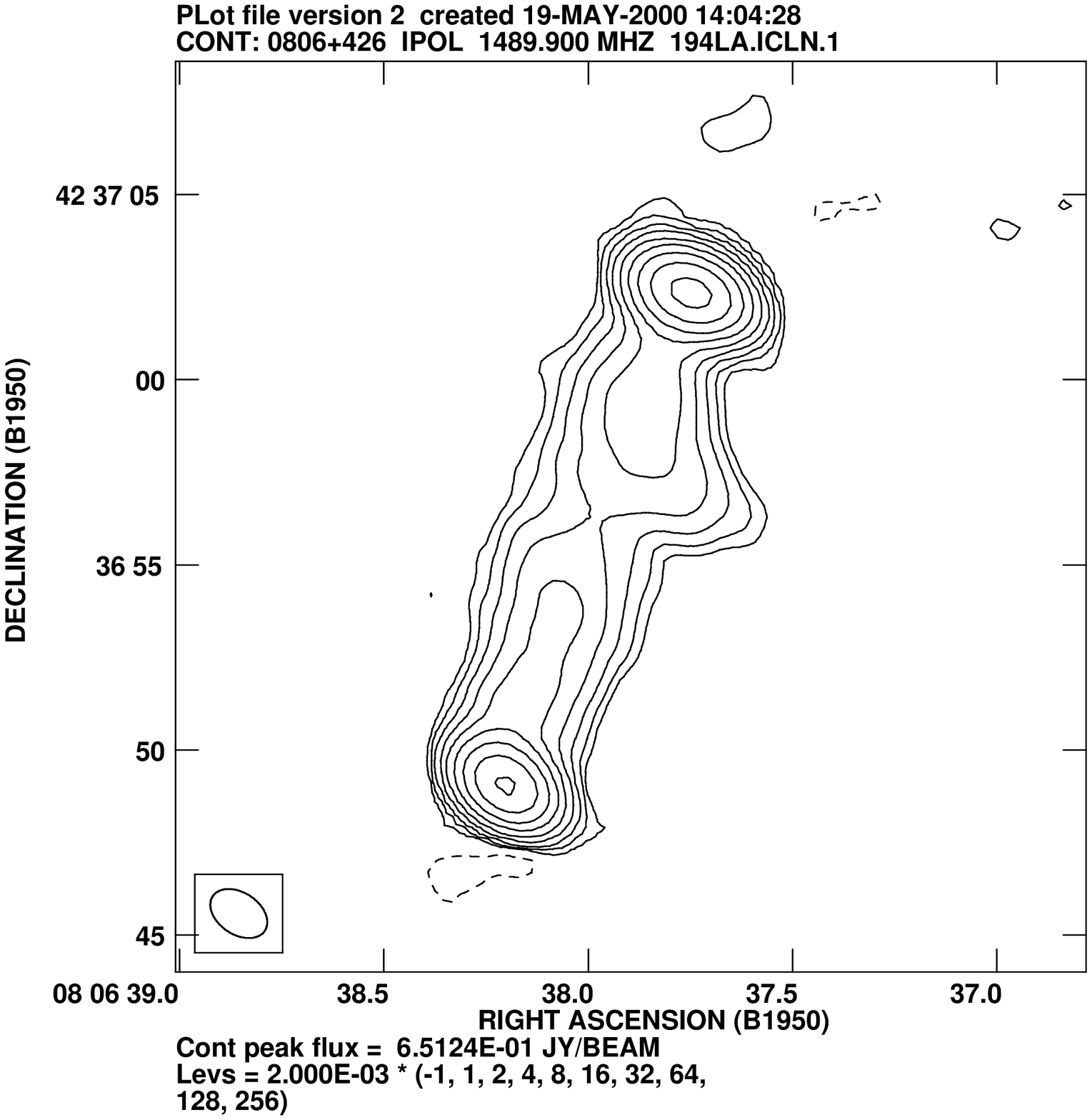}
\centerline{(b)}
\plotone{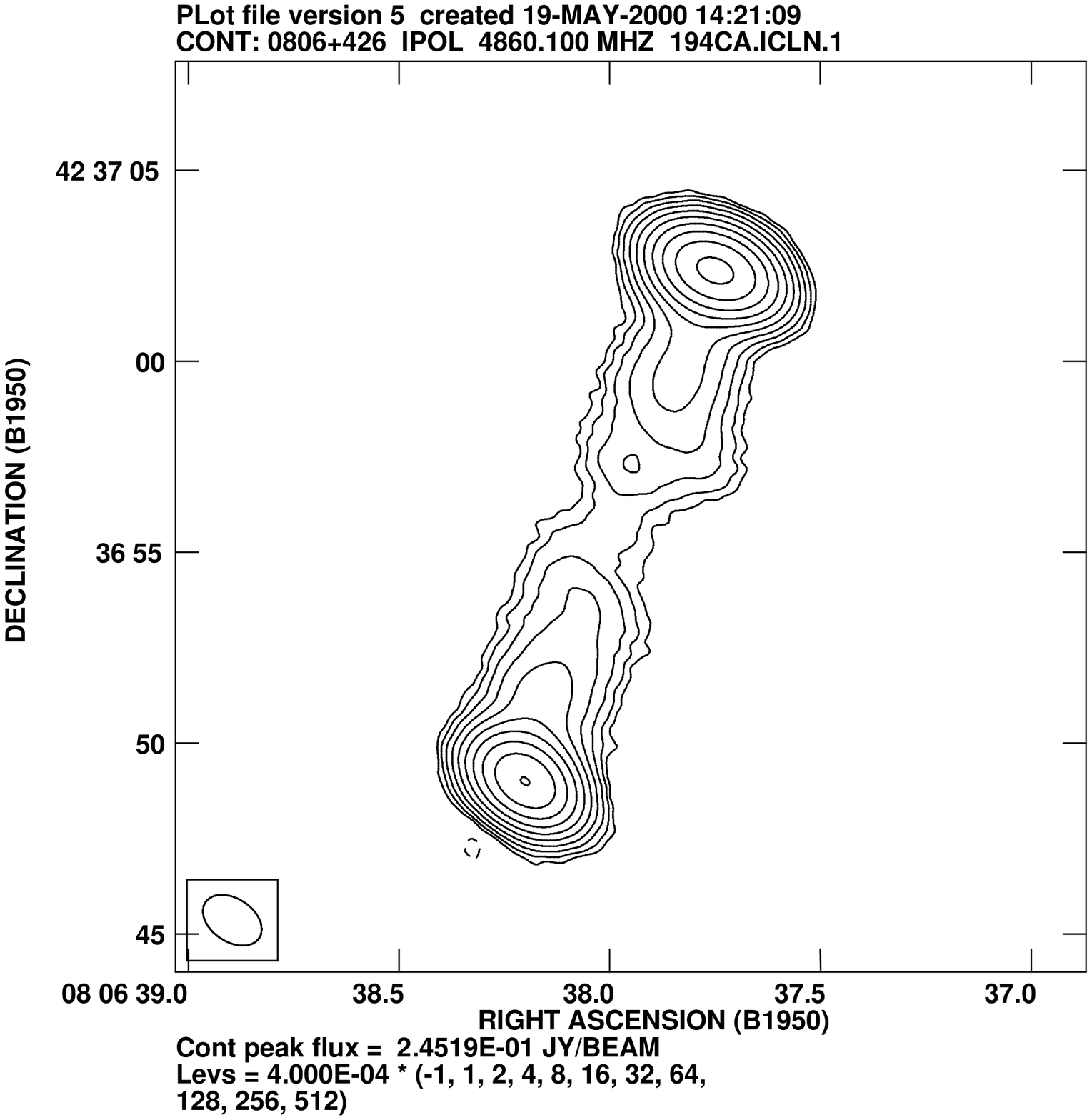}
\caption{Radio maps of 3C 194 from archive data 
in (a) L band and (b) C band. \label{fig:194}}
\end{figure}

\begin{figure}
\epsscale{0.75}
\centerline{(a)}
\plotone{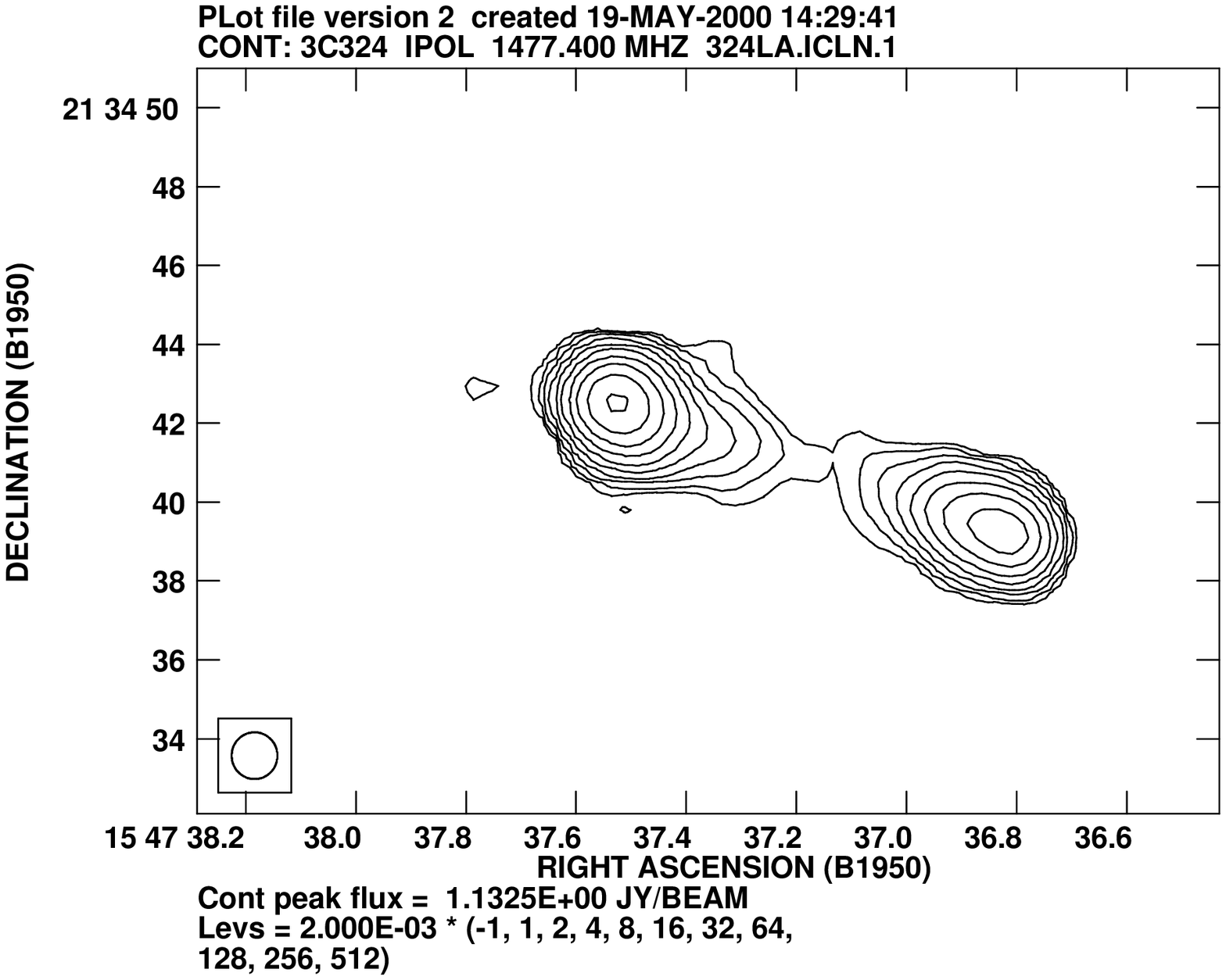}
\centerline{(b)}
\plotone{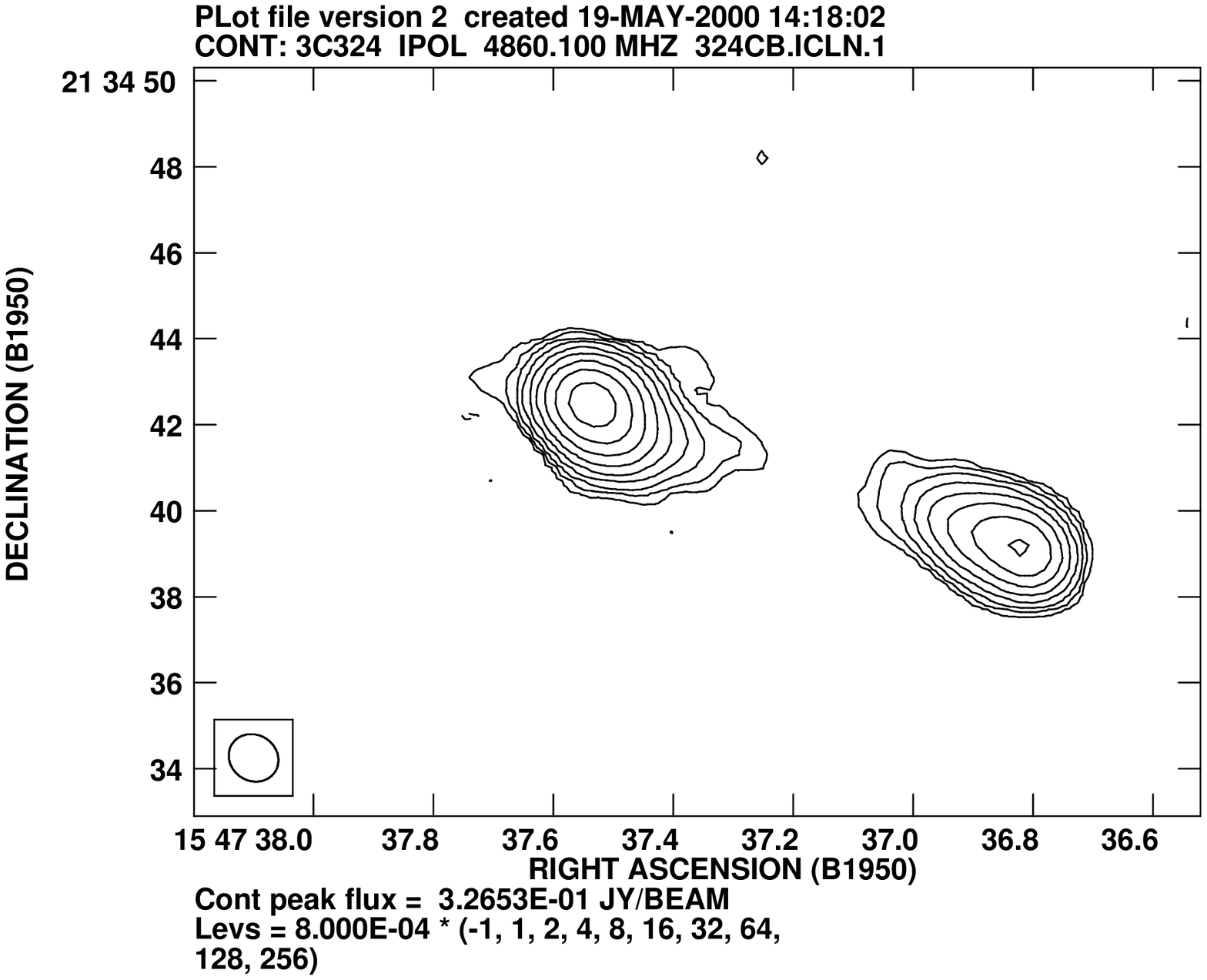}
\caption{Radio maps of 3C 324 from archive data 
in (a) L band and (b) C band. \label{fig:324}}
\end{figure}

\begin{figure}
\epsscale{0.6}
\centerline{(a)}
\plotone{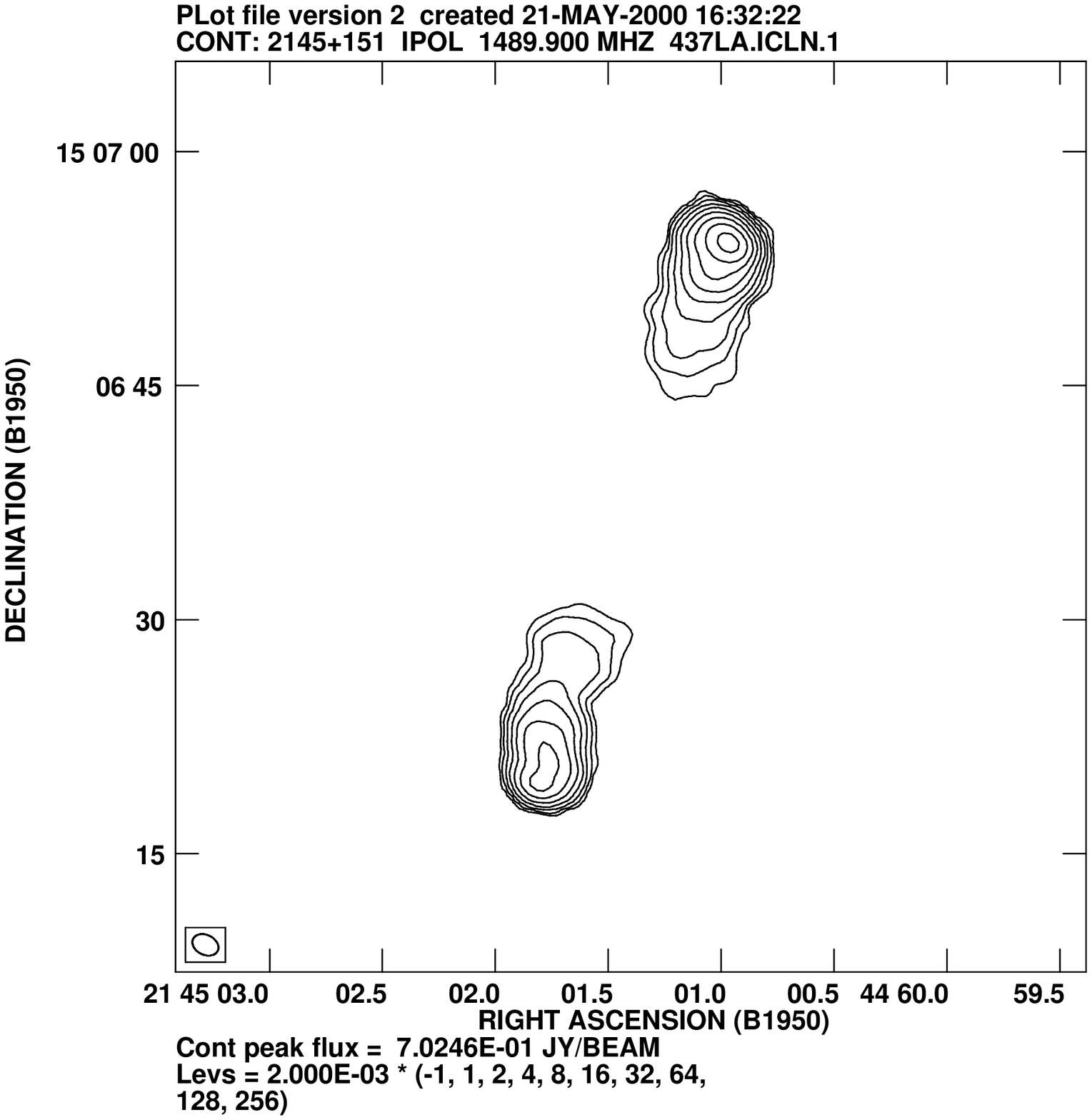}
\centerline{(b)}
\plotone{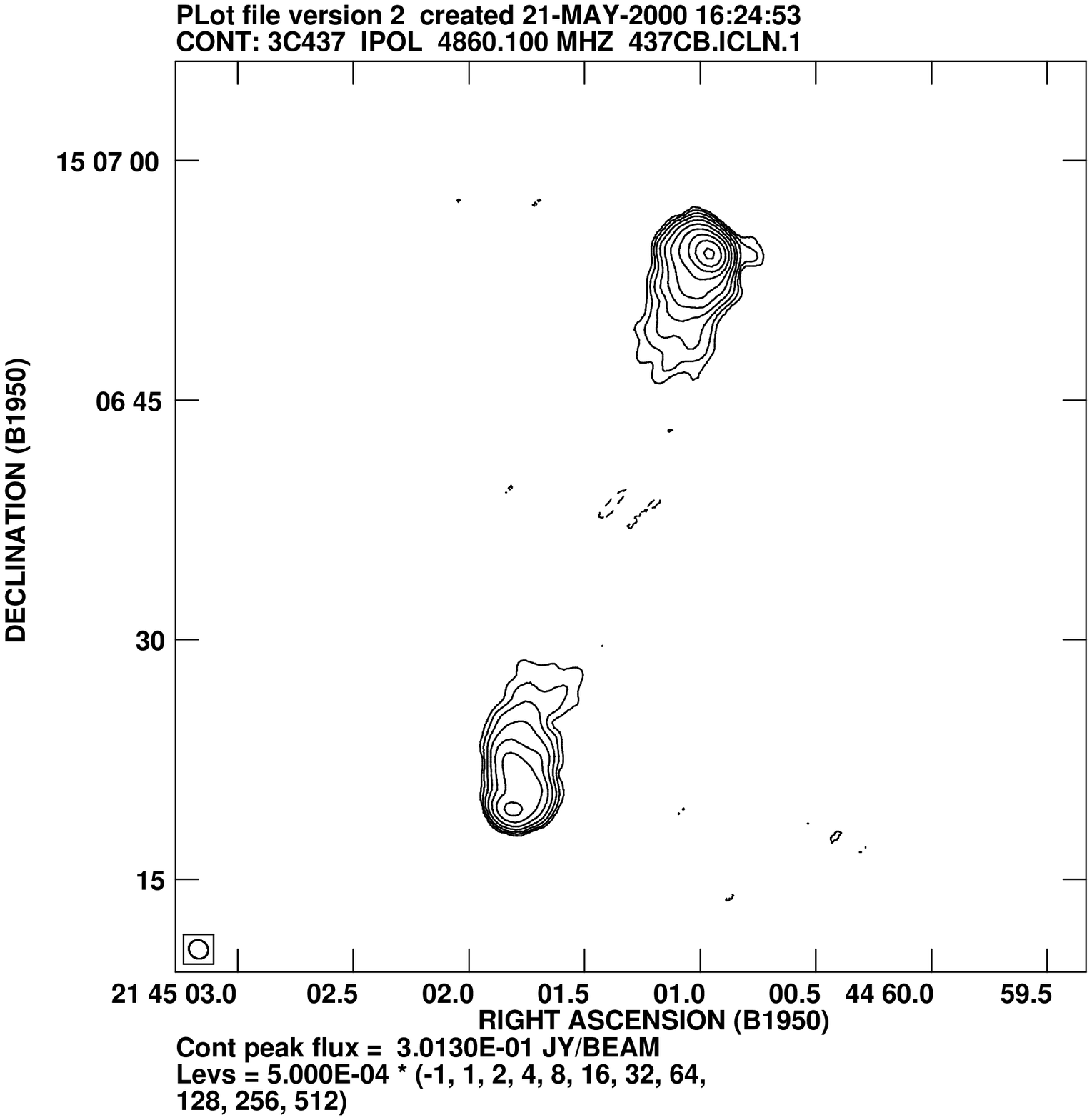}
\caption{Radio maps of 3C 437 from archive data 
in (a) L band and (b) C band. \label{fig:437}}
\end{figure}

Radio intensity 
maps produced from VLA archive data are presented here (Figures
\ref{fig:2441}-\ref{fig:437}).  The 4.872 GHz map of 3C 244.1 was created 
with a restoring beam identical to the synthesized beam
of the 1.411 GHz map of 3C 244.1, and the 4.885 GHz map of 3C 194 
was created with a restoring beam identical to the synthesized beam 
of the 1.465 GHz map of 3C 194.   Thus, radio maps at both frequencies
for a particular radio galaxy
have similar angular resolutions.  Radio maps created with three
of these archive data sets have 
appeared in the literature.  A 1.411 GHz map of
3C 244.1 appeared in Leahy \& Williams (1994), and 4.88 GHz maps of
3C 244.1 and 3C 325 appeared in Fernini, Burns, \& Perley (1997).

The computation of $a_L$, $B_L$, and $v_L$ were performed
here in the same manner as Wellman, Daly, \& Wan (1997a,b).
The deconvolved lobe width, $a_L$, is measured $10 h^{-1}$ kpc
from the hot spot toward the host galaxy.  The lobe magnetic
field strength,
$B_L$, is computed from the deconvolved surface brightness and
bridge width measured $10 h^{-1}$ kpc
from the hot spot toward the host galaxy.  
The lobe propagation velocity, $v_L$,
is computed on the basis of spectral aging along the
imaged portions of the bridge, and the magnetic field used in spectral
aging is computed using values measured $10 h^{-1}$ kpc
and $25 h^{-1}$ kpc from the hotspot.  The errors on each of these
quantities are discussed in \S 5 of WDW97b, \S 5.2 of WDW97a, and 
the appendix of this paper.

All parameters presented here are computed with 
$b=0.25$, which means
magnetic fields are computed to be 0.25 times 
the minimum energy values, and without an $\alpha-z$ correction
which refers to a correction related to 
the observed correlation between spectral index and redshift
(see Wellman 1997;
Wellman, Daly, Wan 1997a,b; Guerra 1998; Guerra \& Daly 1998 for details).
It was found by Guerra \& Daly (1998) that constraints on
cosmological parameters did not depend on these choices, and very
similar results are obtained independent of the value of $b$ and of
whether an $\alpha - z$ correction is applied.  

\subsection{Theory}

The small dispersion in the average size of 3CR radio galaxies at 
a given redshift suggests that, at a given redshift, all of the sources
have a very similar average size.  This size may be estimated by the
average size of the full population of powerful extended radio galaxies
at that redshift, $<D>$, or by the average size of a given source at
that redshift, $D_*$.  If the total time a source produces powerful jets,
with beam power $L_j$, that powers the growth and radio emission of the 
source is $t_*$, then the average source size will be $D_* = v_L ~t_*$, 
assuming the rate of growth of the source, $v_L$, is roughly constant
over the source lifetime.  The source velocity $v_L$, estimated
via synchrotron and inverse Compton aging techniques, increases with 
redshift (Leahy, Muxlow, \& Stephens 1989; Liu, Pooley, \& Riley 1992; 
Daly 1994, 1995; WDW97b), and there is no indication
that the rate of growth of the sources decreases with redshift.  However,
the average size of the full population decreases monotonically with 
redshift for redshifts greater than about 0.5 (see Table 3 in this paper,
and Figure 1 in Guerra \& Daly 1998).  

This means that, for sources of this type, $t_*$ must decrease with redshift.
For the sources studied here, it was shown by WDW97a that the radio power
of a given source is roughly constant over the lifetime of that source, 
making it very unlikely that 
the radio power of a given source decreases with 
time, causing sources at higher redshift to fall below the 
radio flux limit of the survey when they are smaller.   

Thus, since $<D>$ decreases with redshift for $z \ges 0.5$, and 
$D_* = v_L~t_*$ with $v_L$ either increasing with redshift or 
independent of redshift, the data require that $t_*$ decrease with 
redshift.  The rate of growth of the source, $v_L$, depends on one
parameter that is intrinsic to the AGN, the beam power $L_j$, and
on two parameters that are extrinsic to the AGN, the ambient gas
density $n_a$ and the cross sectional area of the radio lobe $\propto a_L^2$:
$v_L \propto (L_j~n_a^{-1}~a_L^{-2})^{(1/3)}$.  The total time for
which the AGN produces a powerful outflow, $t_*$, must depend on 
properties intrinsic to the black hole, and the only intrinsic
parameter that appears in $v_L$ is $L_j$.  Thus, we write 
$t_* \propto L_j^{-\beta/3}$, which defines the one model
parameter $\beta$.  The other intrinsic property of the black
hole that might enter is the total energy available to power
the outflow, $E_*$, but since $E_* = L_j~t_*$, any dependence on
$E_*$ is absorbed into the relation between $t_*$ and $L_j$.  

A power law relation between the total time the AGN produces
powerful jets and the beam power of the jets is not unexpected.  
It means that there is a power law relation between the beam
power $L_j$ and the total energy available initially $E_*$; it
is assumed that the beam power is roughly constant over the 
lifetime of the source, but the value of $L_j$ is set by the 
initial energy available $E_*$.  This 
is reminiscent of 
the power law relation for main sequence stars 
between luminosity and lifetime, or between
total energy available and
luminosity for main sequence stars, and of other power law relations that
arise so frequently in astrophysics.  For example, for a value 
of $\beta$ of 2, the beam power of the jet is related to the total
energy available by the relation $L_j \propto E_*^3$, which is
similar to the relation between luminosity and mass
for main sequence stars: $L \propto M^{3.5}$.  Note, that for
main sequence stars, the luminosity is determined by the
total energy available initially, and remains roughly constant
over the lifetime of the main sequence star.  For a value of 
$\beta$ of 1.5, the relation between beam power and total
energy available is $L_j \propto E_*^2$.  

The excellent fits obtained by comparing $<D>$ with $D_*$ assuming
that $t_* \propto L_j^{-\beta/3}$ supports this choice for the 
parameterization of $t_*$.  As shown in Figure \ref{fig:lin},
if the zero-redshift bin containing Cygnus A is excluded from
the analysis, and the best fit value of $\beta$ and the constant
$C_{ratio}$ (defined below) are determined, 
we can predict the value of $D_*$ for 
Cygnus A, and it matches the value of $D_*$ for Cygnus A to very
high accuracy.  This is quite extraordinary if we consider the 
large drop in $<D>$ that occurs from the first redshift
bin to the zero-redshift bin, which is mirrored by the large
drop in $D_*$ of Cygnus A (see Figures \ref{fig:dave} and 
\ref{fig:dstar}).  

The theory requires that $D_*$ track $<D>$ with redshift, thus it requires
$$<D>/D_* = C_{ratio}~,$$ where $C_{ratio}$ is a constant, independent of 
redshift.  We do NOT require that $<D>/D_* = 1$, but we require that 
this ratio is a constant, $C_{ratio}$.  Thus, the use of this 
equation for cosmology is {\it independent} of any factor
that might be used to normalize the quantity $D_*$.  

The ratio $<D>/D_*$ depends on cosmological parameters and the model
parameter $\beta.$  The way that this ratio depends on coordinate distance 
is given by equation (6) of Guerra \& Daly (1998), and the
full dependence of $D_*$ on observable parameters is given in Appendix
A of this paper.  

The confidence with which cosmological parameters and the model parameter
can be constrained depends on the uncertainty of the ratio $<D>/D_*$.  
This is obtained by combining the uncertainty of $<D>$ and the uncertainty
of $D_*$ in quadrature.  The uncertainty of $<D>$ is given by the dispersion 
in radio source size at a given redshift.  The errors in $<D>$ range 
from 10 \% to 35 \%
(see Table 3).  This dominates the uncertainty of the ratio $<D>/D_*$.  
The uncertainty of $D_*$ depends on the uncertainty of each of the
quantities used to determine $D_*$, and is discussed in detail in Appendix
A of this paper.  

\subsection{Application of the Theory}

In terms of quantities that can be estimated from radio observations,
$L_j \propto v_L~B_L^2~a_L^2$ so
$D_* \propto v_L~L_j^{-\beta/3}$ $\propto v_L^{(1-\beta/3)}~(B_L~a_L)^{-2\beta/3}$ for perfectly symmetric lobes and bridges.  
The three parameters, $a_L$, $B_L$, and $v_L$ 
are computed for the bridge on each side of the six radio galaxies
obtained from the VLA archive, with the exception
of one bridge in 3C 324 which was not imaged along its length sufficiently.
The characteristic core-lobe size, $r_*$, was computed for each bridge using
using the equation described above and in Guerra \&
Daly (1998), and Daly (1994, 1995):
\begin{equation}
r_* \propto \left({1 \over B_L~a_L} \right)^{2 \beta/3}~
v_L^{1 - \beta/3}~~.
\end{equation}
The characteristic 
(lobe-lobe) size, $D_*$, is taken
to be the sum of both $r_*$ (or in the case of 3C 324,
$D_*= 2 r_*$), and is normalized so that $D_*$ of Cygnus A (3C 405), a very
low redshift source in our sample, is equal to the average size of the 
full population of powerful classical double radio galaxies at
very low redshift; this does not impact the determination of 
the model parameter, $\beta$, or constraints on cosmological parameters,
in any way (see \S 2.2).   Constraints on cosmological parameters
determined using this method are independent of the
normalization of $D_*$, as described in \S 2.2. 
As discussed in \S \ref{ssec:cosmo}, the best fit value of the one
model parameter $\beta$ is determined simultaneously with the 
best fit values for the two cosmological parameters that enter,
$\Omega_m$ and $\Omega_{\Lambda}$, the normalized values of the
current values of the mean mass density and the cosmological
constant (see Guerra \& Daly 1998;
Daly, Guerra, \& Wan 1998).  

Table \ref{tab:dstab} presents the six new $D_*$ values
in the last column, assuming the best fit value of $\beta=1.75$
(see \S \ref{ssec:cosmo}).  
Source name and redshift are listed in the first two columns.
The third column lists the redshift bin corresponding to 
the assignments in Guerra \& Daly (1998) and Table \ref{tab:phys}
below. The lobe-lobe angular size of the source is listed in the fourth column,
the fifth and sixth columns list the core-lobe characteristic sizes,
$r_*$, and the characteristic source size is listed in column seven.  

\section{Updated Results}\label{sec:res}

\subsection{Constraints on Cosmological Parameters}\label{ssec:cosmo}

The subsample of sources
with estimates of the characteristic size, $D_*$, has been increased
from 14 to 20, as discussed above in \S \ref{sec:samp}.  
Only sources with physical sizes, defined as the 
projected separation between the radio hot spots, greater than
$20\,h^{-1}$ kpc can be used to determine a characteristic size.  
This is because smaller sources are typically not sufficiently resolved
so that the radio data are useful, and 
the radio lobes of smaller sources are
interacting with the interstellar medium of the host galaxy
rather than the intergalactic/intracluster medium.  
It was decided that this same criterion should be applied to 
the larger comparison sample of powerful 3CR radio galaxies. 
Thus, 
the sample of radio galaxies used to determine
the redshift evolution of the physical size
has been reduced from 82
to 70; twelve radio galaxies were cut because
the physical separation between their lobes was less than
$20\,h^{-1}$ kpc.  
This has a rather small impact on the actual means
and standard deviations of the parent population, as can be seen
by comparing Table \ref{tab:phys} of this paper with Table 1 of Guerra \&
Daly (1998).  
The average lobe-lobe separations as a function of redshift
are listed in Table \ref{tab:phys} for three example choices
of cosmological parameters (matter-dominated, curvature-dominated, and
spatially flat with non-zero cosmological constant). 

To solve simultaneously for the model parameter $\beta$ and the 
cosmological parameters $\Omega_m$ and $\Omega_{\Lambda}$, the
ratio of $D_*$
for each source to $\langle D \rangle$, the average lobe-lobe
size of the parent population in the
corresponding redshift bin,
is fit to a constant, independent of redshift:
$<D>/D_* = C_{ratio}$, where the constant $C_{ratio}$ is allowed
to float when the best fit parameters are determined. 
The value of the constant is a free parameter,
so the normalization of $D_*$ does not affect the
fits in any way.
  
\begin{figure}
\epsscale{1}
\plotone{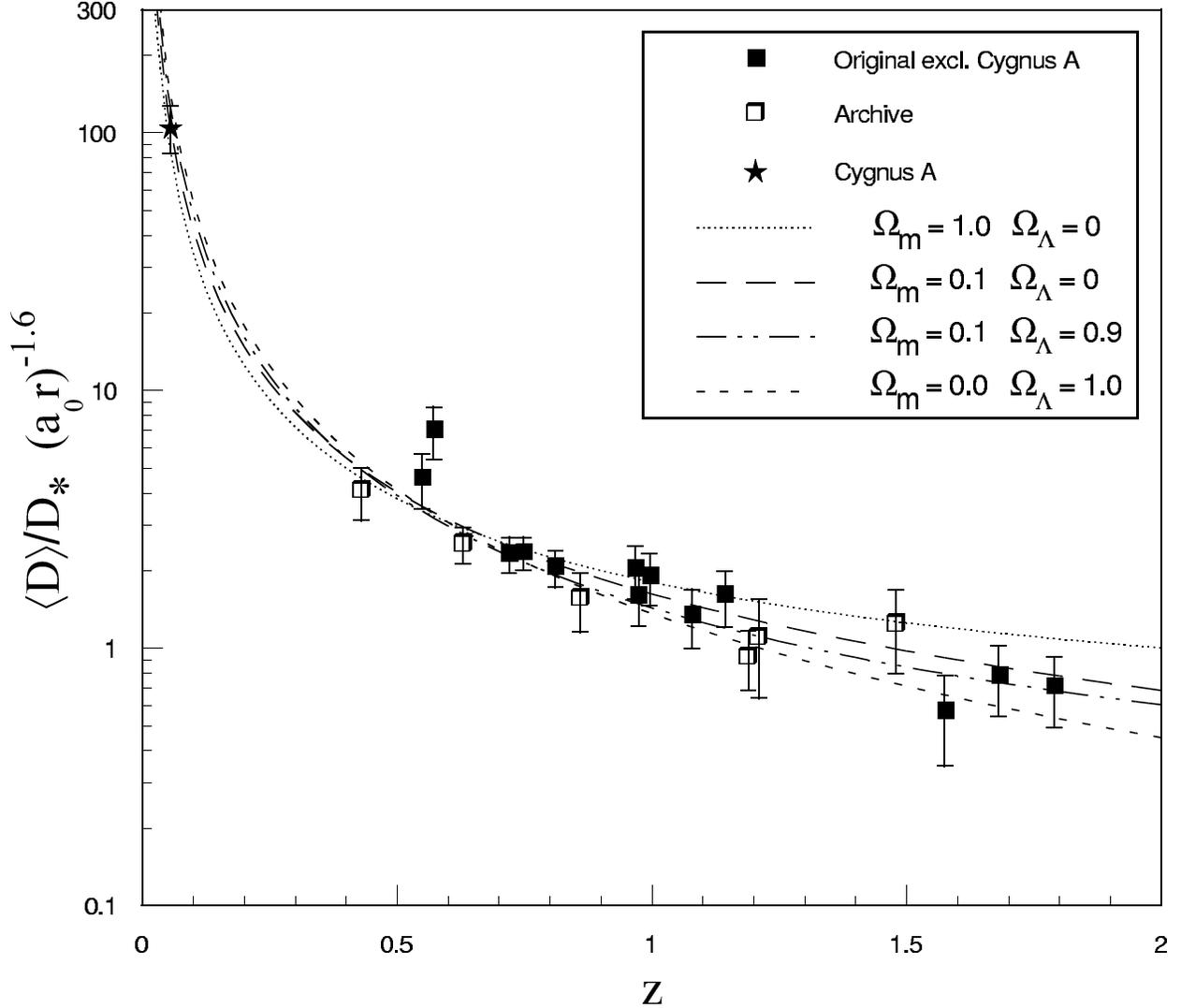}
\caption{The quantity $(\langle D \rangle / D_*) (a_o r)^{-1.6}$,
computed assuming $\beta=1.75$ and $b=0.25$. 
This measured quantity is independent $\Omega_m$ and
$\Omega_{\Lambda}$.
Open squares denote the six new
data points from the VLA archive and Cygnus A is denoted by the star symbol.
For different choices of $\Omega_m$ and $\Omega_{\Lambda}$, 
the best-fit redshift evolution of 
$(a_o r)^{-1.6}$
{\bf excluding Cygnus A} is shown.  Note that the fits are not 
normalized to Cygnus A, but are consistent with Cygnus A.
\label{fig:lin}}
\end{figure}

\begin{figure}
\plotone{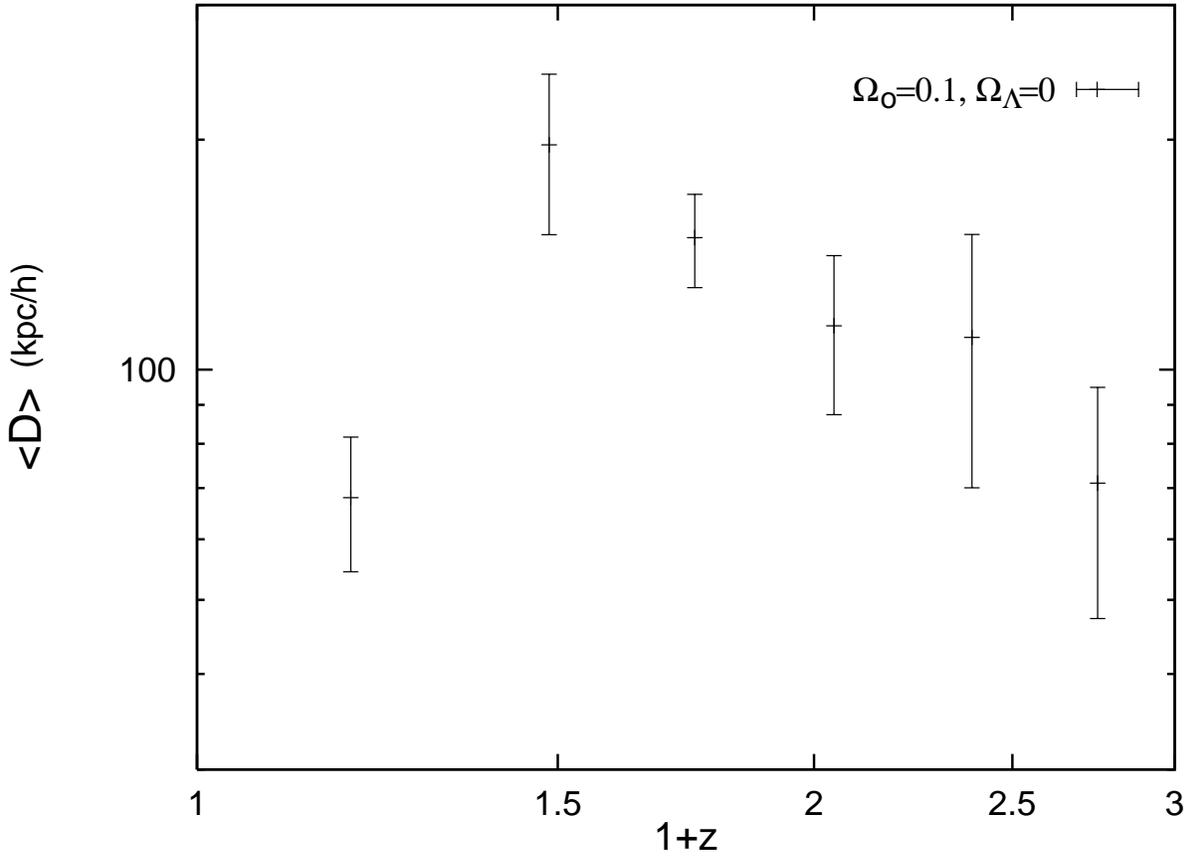}
\caption{The average lobe-lobe size $\langle D \rangle$ of
powerful extended 3CR radio galaxies
computed assuming $\Omega_m=0.1$ and
$\Omega_{\Lambda}=0$.
\label{fig:dave}}
\end{figure}

\begin{figure}
\plotone{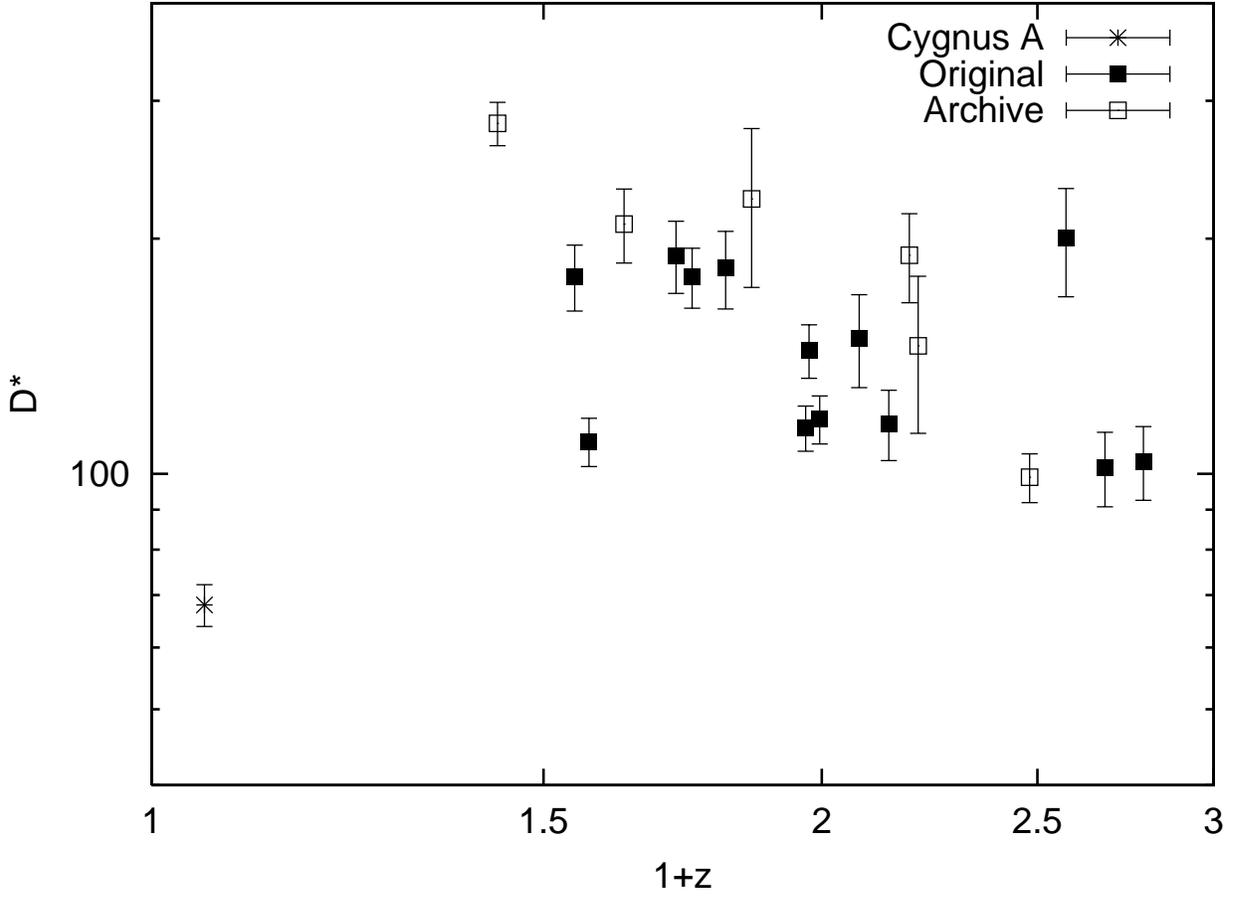}
\caption{The quantity $D_*$
computed assuming $\beta=1.75$, $b=0.25$, $\Omega_m=0.1$, and
$\Omega_{\Lambda}=0$.
Open squares denote the six new
data points from the VLA archive and Cygnus A is denoted by the star symbol.
\label{fig:dstar}}
\end{figure}

Figure \ref{fig:lin} illustrates the cosmological
dependence of $\langle D \rangle / D_*$ on the coordinate distance
$(a_o r)$. As described below, the best fit value of
$\beta$ obtained here is $\beta = 1.75$; for this value
of $\beta$, $\langle D \rangle / D_*$ is
proportional to $(a_or)^{1.6}$.  Thus
$(\langle D \rangle / D_*)(a_or)^{-1.6}$ is independent of
cosmological parameters.  The data can be compared to
several different sets of cosmological parameters on a single
figure by plotting $(\langle D \rangle / D_*)(a_or)^{-1.6}$ for
each data point and comparing this with $(a_or)^{-1.6}$ curves
obtained for different sets of cosmological parameters, as is
shown in Figure \ref{fig:lin}.  Note that, for the fits shown,
the lowest redshift bin, which includes Cygnus A, was excluded
from the analysis.  Even so, the lines all go directly through
this point (denoted by a star) indicating the predictive power
of this model.  

The hypothesis is that, for the 
correct choice of cosmological parameters, $\langle D \rangle /
D_* = C_{ratio}$ = constant, so that the values
of $(\langle D \rangle / D_*)(a_or)^{-1.6}$ for 
all 20 radio galaxies should follow a curve that, at each z, tracks 
the curve $(a_or)^{-1.6}$ obtained for that particular choice of
cosmological parameters.  Figure \ref{fig:lin}, shows $(\langle D \rangle /
D_*)(a_or)^{-1.6}$ for the the six new
points and 14 original points as a function of z
(the six new points are denoted by open squares).  
Also drawn on this figure are the best-fit curves of $(a_or)^{-1.6}$ 
obtained for specified values of cosmological parameters and 
excluding Cygnus A from the fits.  {\bf In
this figure, all of the curves pass through Cygnus A, though
this is not required when we actually solve for best fitting
cosmological parameters.}  Including Cygnus A has a negligible effect on the
normalization of these curves and a small effect on cosmological constraints.
It is clear that curves obtained for a 
low density universe describe the data points quite well, and the 
curve describing a universe with $\Omega_m = 1$ does not follow 
the data points.  

\begin{figure}
\epsscale{0.55}
\centerline{(a)}
\plotone{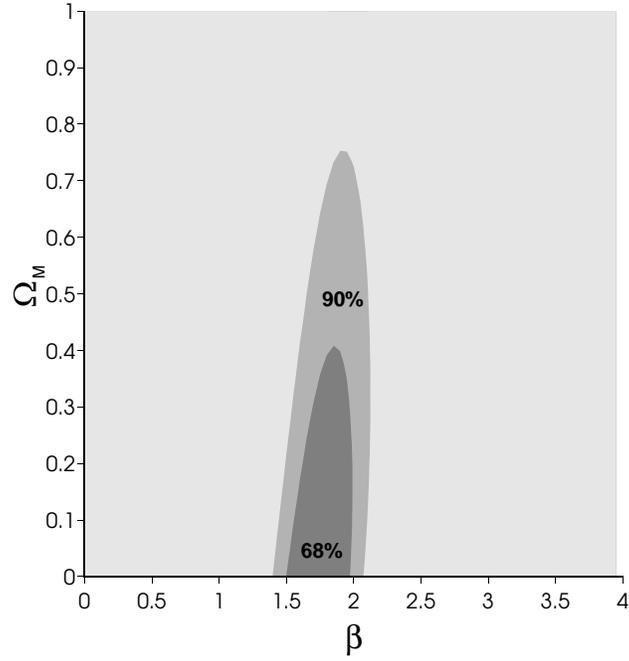}
\centerline{(b)}
\plotone{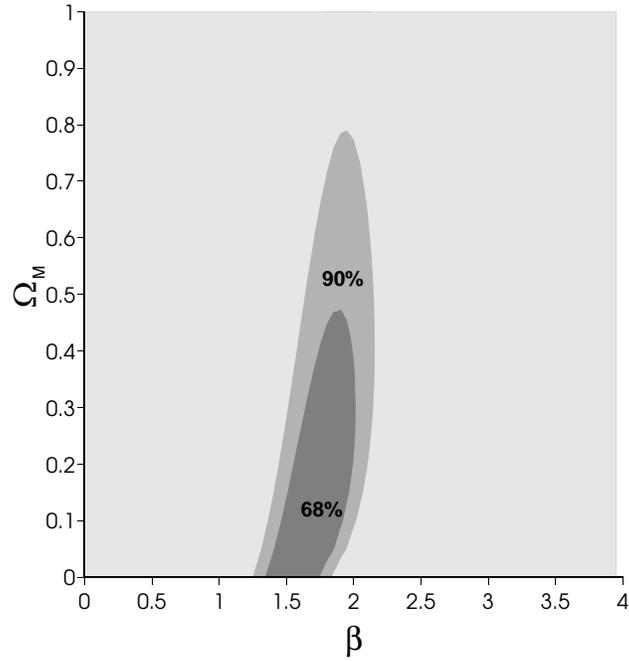}
\caption{The 68\% and 90\% confidence intervals for 
$\beta$ and $\Omega_m$, where (a) $\Omega_{\Lambda}$=0 (no cosmological
constant)and (b) $\Omega_{\Lambda}=1-\Omega_m$ (spatially flat universe).
\label{fig:ombet}}
\end{figure}

For all 20 sources, the chi-squared for fitting
the ratio to a constant is computed for relevant values of 
$\beta$, $\Omega_m$, and $\Omega_{\Lambda}$.  It is found 
that the best fit value of $\beta$ is $\beta = 1.75 \pm 0.25$. 
This result is insensitive to the choice of $\Omega_m$
and $\Omega_{\Lambda}$, and there appears to be no significant
covariance between $\beta$ and cosmological parameters
(see Figures \ref{fig:ombet}a and \ref{fig:ombet}b). 

\begin{figure}
\epsscale{0.75}
\plotone{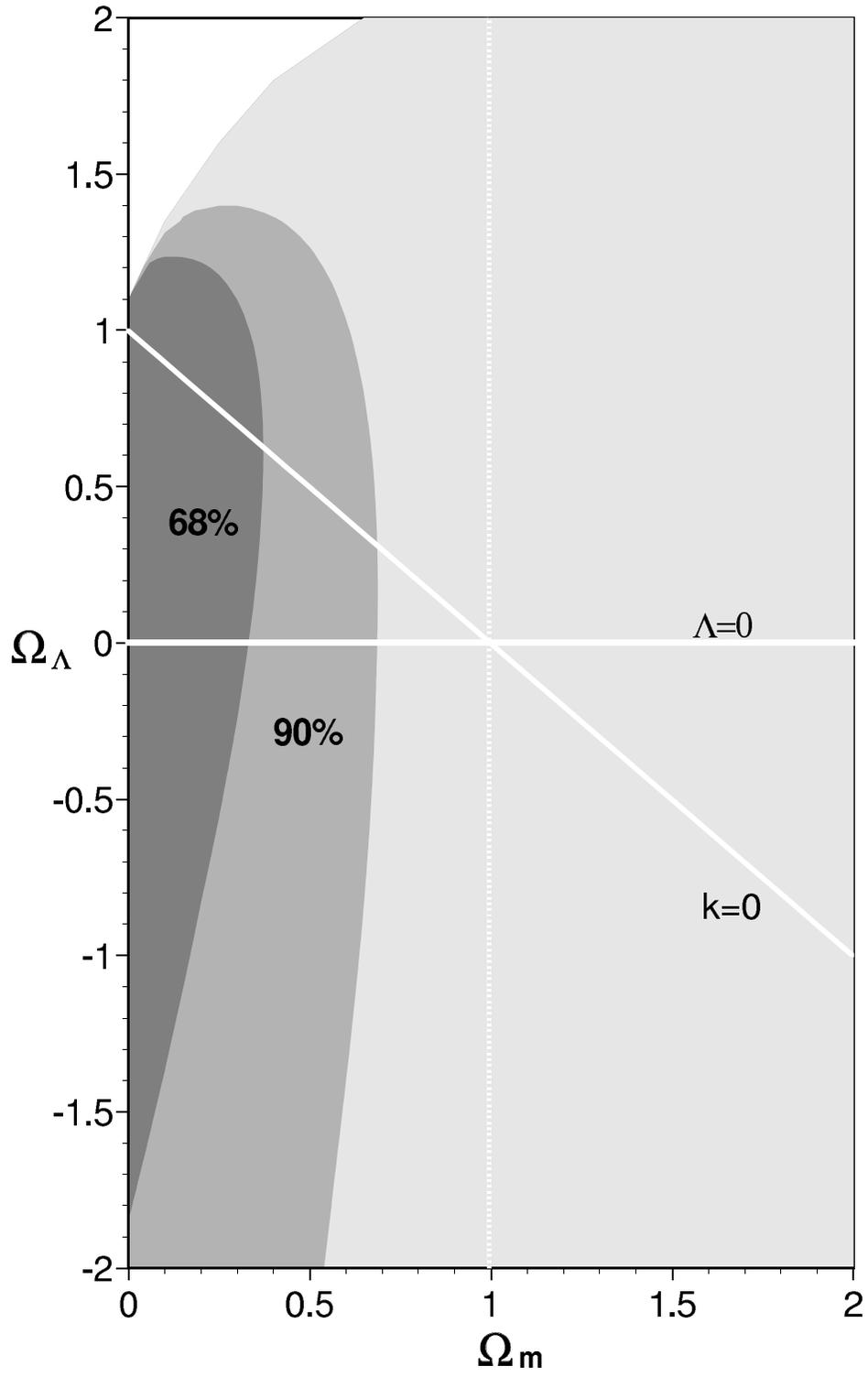}
\caption{The 68\% and 90\% confidence intervals for ranges of both
$\Omega_m$ and $\Omega_{\Lambda}$, independent of $\beta$. (Two-dimensional)
\label{fig:2d}}
\end{figure}

\begin{figure}
\plotone{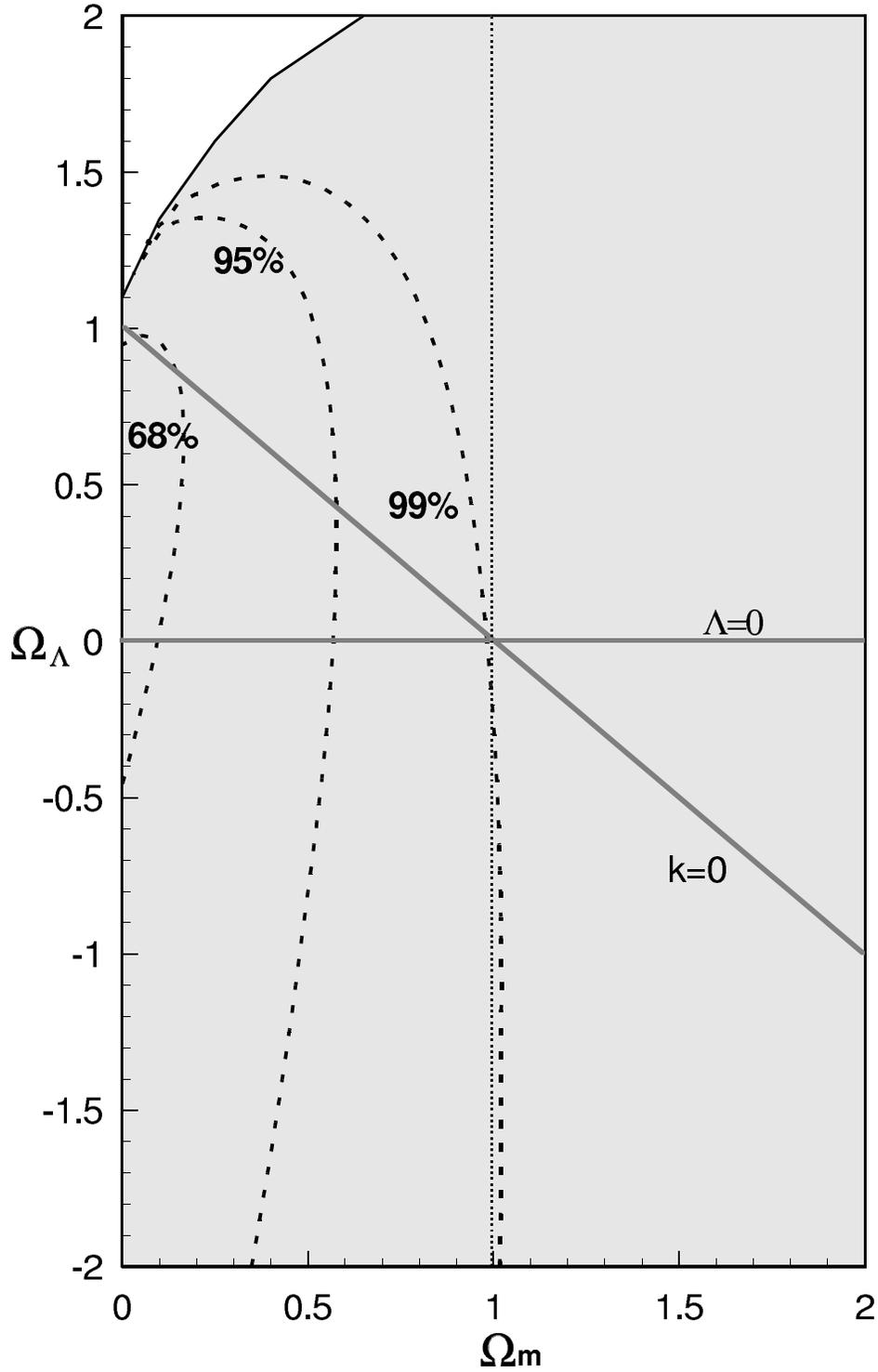}
\caption{The projections of the
68\%, 95\%, and 99\% confidence intervals
onto either axis ($\Omega_m$ or $\Omega_{\Lambda}$) 
indicates the probability associated with the
range in that one parameter, independent of
all other parameter choices. (One-dimensional) \label{fig:1d}}
\end{figure}

The confidence contours in the $\Omega_m$ - $\Omega_{\Lambda}$
plane are shown in 
Figures \ref{fig:2d} and \ref{fig:1d}.
The probability
associated with a given range of
$\Omega_m$ and $\Omega_{\Lambda}$ independent of $\beta$ is
shown in Figure \ref{fig:2d} (referred to as two-dimensional
confidence intervals). 
In Figure \ref{fig:1d}, the projection of a confidence interval
onto either axis ($\Omega_m$ or $\Omega_{\Lambda}$) 
indicates the probability associated with a given
range of that one parameter, independent of
all other parameter choices (referred to as one-dimensional confidence
intervals).
Both figures illustrate how this method and the data
are most consistent with a low density universe; $\Omega_m \les 0.15$
with 68\% confidence, $\Omega_m \les 0.5$ with 90\% confidence,
and $\Omega_m \les 1.0$ with 99\% confidence.
The constraints on $\Omega_{\Lambda}$ are not as strong, and values
of $\Omega_{\Lambda}$ from zero to unity are consistent
with the data.

\begin{figure}
\plotone{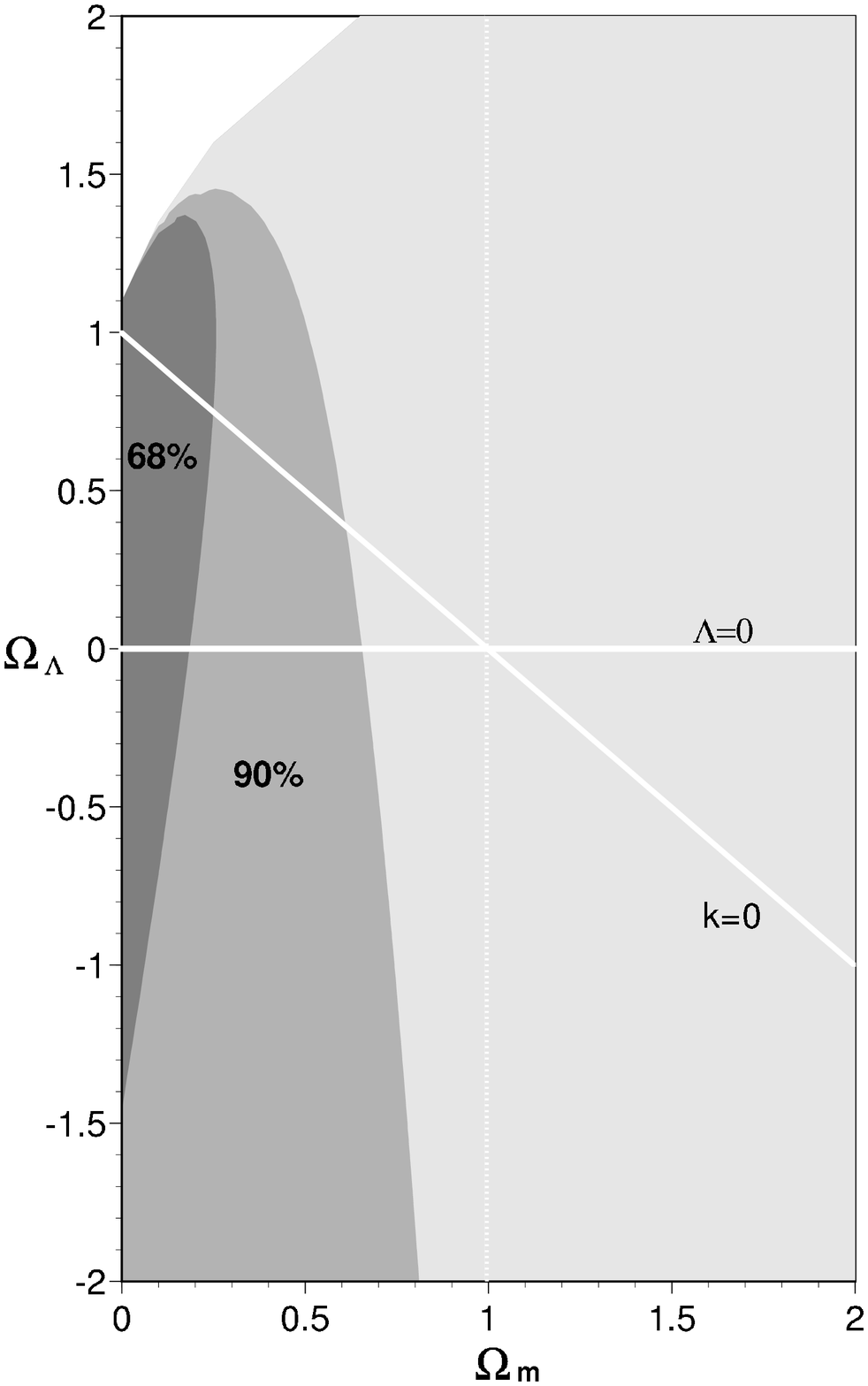}
\caption{The same as Figure \ref{fig:2d}, but excluding Cygnus A from the fit.
\label{fig:cyg2d}}
\end{figure}

\begin{figure}
\plotone{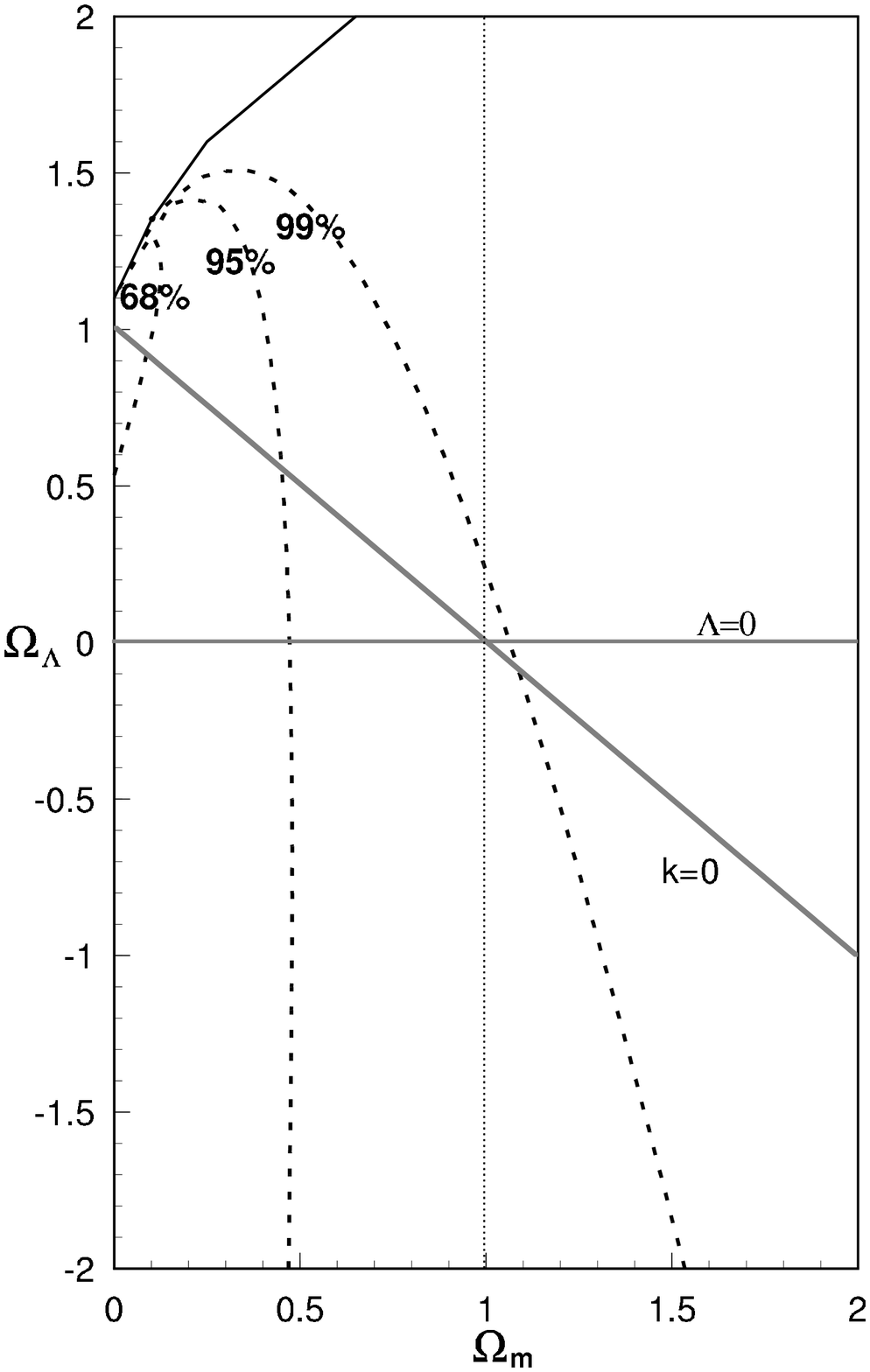}
\caption{The same as Figure \ref{fig:1d}, but excluding Cygnus A from the fit.
\label{fig:cyg1d}}
\end{figure}

The best fit value of $\beta = 1.75 \pm 0.25$ is consistent with 
the previous estimates of $\beta \simeq 1.5 \pm 0.5$ (Daly 1994), and
$\beta \simeq 2.1 \pm 0.6$ (Guerra \& Daly 1996, 1998), but with
significantly reduced uncertainties. Similarly, the constraints on 
cosmological parameters are consistent with previous estimates, 
(Daly 1994; Guerra \& Daly 1996, 1998; Guerra 1997; Daly, Guerra, \& Wan 1998) 
but with smaller error bars.  
It is apparent in Figures \ref{fig:2d} and \ref{fig:1d} that 
these data and method
strongly favor a low density universe; a universe where 
$\Omega_m = 1$ is ruled out with 99.0 \% confidence
independent of $\Omega_{\Lambda}$ and $\beta$ . 
This will be discussed in more
detail in \S \ref{sec:disc}.  To further illustrate the insensitivity of 
cosmological constraints to
including Cygnus A, Figures \ref{fig:cyg2d} and \ref{fig:cyg1d} show
the confidence intervals when Cygnus A is excluded from the fits.

Figures \ref{fig:cyg2d} and \ref{fig:cyg1d}
also illustrate that that data
are most consistent with a low density universe.
Fits excluding Cygnus A rule out an $\Omega_m = 1$ universe
at 97.5\% confidence.
These results are quite similar to those obtained
when Cygnus A is included.  This further illustrates that the
method does not
rely upon a low-redshift normalization.

\subsection{Ambient Gas Densities and Beam Powers}\label{ssec:other}

Daly (1994, 1995, 2000), following the work of Perley \& Taylor (1991), 
Carilli et. al (1991), and Rawlings \& Saunders (1991), 
showed that the radio properties of a powerful
extended radio source could be used to estimate the beam power,
$L_j$, and density of the gas in the vicinity of the source,
$n_a$.  The method 
is described in detail by Wellman (1997), Wellman, Daly, \& Wan 
(1997a), Wan (1998),
Wan, Daly, \& Guerra (1998), and Daly (2000).  Values for the ambient
gas density and beam power for the 14 sources in the original
sample are described in these papers.  
The basic equations are:
\begin{equation}
L_j \propto a_L^2~P_L~v_L~~,
\end{equation} 
and
\begin{equation}
n_a \propto {P_L \over v_L^2}~~; 
\end{equation}
where $P_L$ is the lobe pressure (see equation A2), and 
the normalizations are
given in the references listed above. The values obtained for the six new
radio galaxies in our sample are listed in Table \ref{tab:natab}.  Also
listed
in Table \ref{tab:natab} are all the input parameters 
used to compute $L_j$, $n_a$, and $D_*$.

\section{Discussion}\label{sec:disc} 

A parent population of 70 powerful extended classical double
radio galaxies with redshifts between zero and two was used to 
define the evolution of the mean or characteristic size of these
sources as a function of redshift; these sources are all
Type 1 FRII sources (Leahy \& Williams 1984), also referred
to as FRIIb sources (Daly 2000).  An independent estimate of
the mean or characteristic size of a given source was possible
for a subset of 20 of these radio galaxies for which extensive
multiple frequency radio data was available.  Requiring that the two 
measures of the mean source size have the same redshift behavior
allows a simultaneous determination of three parameters: the one
model parameter $\beta$, and the two cosmological parameters
$\Omega_m$ and $\Omega_{\Lambda}$ (assuming that the only
significant cosmological parameters today are the mean mass
density, a cosmological constant, and space curvature).  

The method was applied to this data set, and interesting new constraints are
presented.  It is found that the model parameter is very tightly
constrained to be $\beta = 1.75 \pm 0.25$ 
(see Figure \ref{fig:ombet}), consistent with 
previous estimates, and that this model parameter is independent
of cosmological parameters.  For a value of $\beta = 1.75$, the 
characteristic source size is $D_* \propto
(B_La_L)^{-7/6}~{v_L}^{5/12}$, which indicates that it is necessary
to have multiple-frequency radio data in order to estimate
$D_*$, owing to its $v_L$ dependence.  

The data strongly favor a low density universe; a universe with 
$\Omega_m = 1$ is ruled out with 99\% confidence, independent
of the value of $\Omega_{\Lambda}$ or $\beta$.  Either space
curvature or a cosmological constant, or both, are allowed.  
The main conclusion is that $\Omega_m$ is low, but, at this
point, the method and data do not allow a discrimination between 
whether space curvature or a cosmological constant is important 
at the present epoch.  

It is interesting to note that the lowest
reduced chi-squared obtained is 0.96 for $\Omega_m \simeq -0.25$
and $\Omega_{\Lambda} \simeq 0$.
This value is slightly
greater and closer to unity than the minimum reduced chi-squared
obtained by Guerra \& Daly (1998), which indicates that
this sample of 20 sources has a reasonable distribution
around any model predictions.  This convergence
to unity with increasing sample size suggests
that this method and its statistics are reliable.

The best fit for cosmological parameters in the
physically relevant half-plane of $\Omega_m \ges 0$
is $\Omega_m = 0$ and $\Omega_{\Lambda} = 0.45$ with a
reduced chi-squared of 0.98 (also close to unity).
However,
Figures \ref{fig:2d} and \ref{fig:1d} clearly show that
our results are still consistent with $\Omega_{\Lambda} = 0$.

The radio data also allow a determination of the density of the 
ambient gas in the vicinity of each radio source, and the beam
power of each source; the values of these quantities are
presented.  The sources lie in high-density gaseous environments
like those found in low-redshift clusters of galaxies.  Typical
beam powers for the sources are $\sim 10^{45} \hbox{ erg
s}^{-1}$.  

\acknowledgments

Special thanks go to Rick Perley and Miller Goss for their aid in extracting 
data from the VLA archive.
The National Radio Astronomy Observatory is a facility of the 
National Science Foundation operated under 
cooperative agreement by Associated Universities, Inc. 
The authors would also like to thank Katherine Blundell, 
Chris Carilli, Paddy Leahy, Wil van Breugel,
and Dave Wilkinson for helpful discussions.  
Special
thanks are extended to the referee for providing valuable feedback
and suggestions that led to significant improvement of the paper.    
This work was supported in part by the U.S. National Science
Foundation and the College of Liberal Arts and Sciences at Rowan University. 

\appendix

\section{Error Analysis}

Errors on $D_*$ and errors on $<D>$ both contribute to the confidence contours 
relevant for the cosmological parameters determined here.  The errors on $<D>$ 
are purely statistical, and arise from the dispersion in source size of the 
parent population, and at the present time dominate the total uncertainty of 
the quantity $<D>/D_*$.  The errors on $D_*$ depend on the uncertainties of 
quantities used to determine $D_*$; the way that the error on $D_*$ is 
computed is presented here.  

The quantity 
\begin{equation}
r_* = v_L~ t_* \propto v_L~ L_j^{-\beta/3}
\propto v_L \left(P_L~v_L~a_L^2\right)^{-\beta/3}
\propto v_L^{1-\beta/3}~a_L^{-2\beta/3}~P_L^{-\beta/3}.
\end{equation}  
The lobe 
pressure 
\begin{equation}
P_L \propto \left(\frac{4}{3}~b^{-1.5} + b^2\right)~B_{min}^2, 
\end{equation}
where $B_{min}$ is 
the minimum energy magnetic field, $B_{min} \propto S_{\nu}^{2/7}~a_L^{-2/7}$, 
and is determined  using parameters obtained 10 $h^{-1}$ kpc behind the 
hotspot: the radio surface brightness at this location is
$S_{\nu,10}$ and the lobe half-width at this location is $a_{L,10}$.
The radio surface brightness and lobe half-width 25 $h^{-1}$ kpc
behind the hotspot are denoted $S_{\nu,25}$ and $a_{L,25}$ respectively.    
The lobe propagation velocity $v_L$ is estimated using a standard 
synchrotron and inverse Compton aging model in which the time, $t$, required 
for the source to grow a size $\Delta x$, is used to estimate the lobe 
propagating velocity $v_L = \Delta x/t$.  The time 
\begin{equation}
t \propto \frac{B_{av}^{1/2}}{\nu_T^{1/2}\left(B_{av}^2+B_{MB}^2\right)}
\end{equation} 
(see, for example, 
Wan \& Daly 1998, \S 3) where $\nu_T$ is the break frequency, $B_{av}$ is the 
average magnetic field in the radio bridge, taken to be $B_{av} = 
\sqrt{B_{10}B_{25}}$ (WDW97b), $B_{10}$ is the magnetic field strength 10 
$h^{-1}$ kpc behind the hotspot, $B_{25}$ is the magnetic field strength 
25 $h^{-1}$ kpc behind the hotspot, and $B_{MB}$ is the term that describes 
inverse Compton cooling of relativistic electrons by scattering with microwave 
background photons, and is obtained by equating the energy density of the 
microwave background at a given redshift with $B_{MB}^2/(8 \pi)$.  

In terms of these parameters, 
\begin{eqnarray}
r_* &
\propto & \left( \frac{4}{3}~ b^{-1.5} +b^2 \right) ^{-\beta/3}~B_{10}^{-2\beta/3}~a_L^{-2\beta/3} 
~\Delta x^{(1-\beta/3)} ~\nu_T^{\frac{1}{2}(1-\beta/3)} \\
& & \times \left(B_{av}^2+B_{MB}^2\right) ^{1-\beta/3}
~B_{10}^{-\frac{1}{4}(1-\beta/3)}~B_{25}^{-\frac{1}{4}(1-\beta/3)}.\nonumber
\end{eqnarray} 
Given that the 
magnetic field strength is parameterized by $B=b~B_{min}$ at any given 
location, we obtain the following expression for $r_*$, which is relevant 
for a determination of the error: 
\begin{eqnarray}
r_* & \propto & 
\left(\frac{4}{3}b^{-1.5} + b^2 \right)^{-\beta/3} b^{-\frac{1}{2}(1+\beta)}~\Delta x^{(1-\beta/3)}~{\nu_T}^{\frac{1}{2}(1-\beta/3)}~
{B_{25}}^{\frac{1}{4}(\beta/3 - 1)}
~{a_{L,10}}^{-\frac{1}{2}(\beta - 1/7)}~ \\
& & \times {S_{\nu,10}}^{-(\beta/6 + 1/14)} ~
\left(b^2~{S_{\nu,10}}^{2/7}~
{S_{\nu,25}}^{2/7}~{a_{L,10}}^{-2/7}~{a_{L,25}}^{-2/7}+
{B_{MB}}^2\right)^{(1-\beta/3)}. \nonumber
\end{eqnarray}  
The total error on $\delta r_*/r_*$ is found by 
taking the partial derivative of $r_*$ with respect to a given variable, for 
each variable, adding these terms in quadrature, and taking the square-root.  
We obtain
\begin{equation}
\left(\frac{\delta r_*}{r_*}\right)_{\nu_T} = \frac{1}{2}~(1-\beta/3)
\left(\frac{\delta \nu_T}{\nu_T}\right)
\end{equation}
\begin{equation}
\left(\frac{\delta r_*}{r_*}\right)_{\Delta x} = (1-\beta/3)
\left(\frac{\delta\Delta x }{\Delta x}\right)
\end{equation}
\begin{equation}
\left(\frac{\delta r_*}{r_*}\right)_{B_{25}}=\left[-\frac{1}{4}(1-\beta/3) + 
(1-\beta/3) \left( \frac{B_{10}B_{25}}
{B_{10}B_{25}+{B_{MB}}^2} \right) \right]
\left(\frac{\delta B_{25}}{B_{25}}\right)
\end{equation}
\begin{equation}
\left(\frac{\delta r_*}{r_*}\right)_{a_{L,10}} = 
\left[\frac{1}{2}(1/7 - \beta) - \frac{2}{7}(1-\beta/3) 
\left( \frac{B_{10}B_{25}}
{B_{10}B_{25}+{B_{MB}}^2} \right) \right]
\left(\frac{\delta a_{L,10} }{a_{L,10}}\right)
\end{equation}
\begin{equation}
\left(\frac{\delta r_*}{r_*}\right)_{S_{\nu,10}}= \left[ -(\beta/6+1/14)
+\frac{2}{7}(1-\beta/3)
\left( \frac{B_{10}B_{25}}
{B_{10}B_{25}+{B_{MB}}^2} \right) \right]
\left(\frac{\delta S_{\nu,10} }{S_{\nu,10}}\right)
\end{equation}
\begin{equation}
\left(\frac{\delta r_*}{r_*}\right)_{b}=
\left[ \frac{ 2\beta}{3} \left(\frac{ 1-b^{3.5}}{4/3+b^{3.5}}\right)-
\frac{1}{2}(1+\beta) + (1-\beta/3) \left(
\frac{ 2 B_{10}B_{25}}{B_{10}B_{25}+{B_{MB}}^2} \right)
\right]
\left(\frac{\delta b}{b}\right)
\end{equation}

The final uncertainty of $r_*$ divided by $r_*$, $\delta r_*/r_*$ 
is obtained by adding each of the terms listed above in quadrature,
and taking the square root.  
As mentioned above, this is typically
much less than the uncertainty of $<D>$, so the confidence level
with which cosmological parameters are determined is primarily controlled
by the uncertainty in $<D>$ (see Table 3).  

The uncertainties on each of the quantities listed above are described
in  section 5 of WDW97b, and section 5.2 of
WDW97a.  
The uncertainty in $\nu_T$ is typically (10 to 20) \%.  
The uncertainty on $\Delta x$ is typically (6 to 20) \%. 
The uncertainty on
$B_{25}$ is typically 6 \%.  The uncertainty on $a_{L,10}$ is typically
(6 to 20)\%; this term (et. A9) is generally the largest contribution to 
the error in $r_*$.  The uncertainty on $S_{\nu,10}$ is typically
20 \%, and the uncertainty on b is about 15 \% for b = 0.25, and is about
10 \% for b = 1 (see Wellman, Daly,\& Wan 1997b, section 7.4).

For a given value of $\beta$, 
equations A6 through A11 may be written as a number times
the fractional error on the relevant quantity.  
Some of the equations involve the 
fraction $f= \left(
\frac{ B_{10}B_{25}}{B_{10}B_{25}+{B_{MB}}^2} \right)$, 
which is always less than or equal to one.  To get a 
sense of which terms contribute to the total error 
in $r_*$, equations A6 through A11 are rewritten 
assuming a value of $\beta$ of 1.75.  

\begin{equation}
\left(\frac{\delta r_*}{r_*}\right)_{\nu_T} \simeq 0.21
\left(\frac{\delta \nu_T}{\nu_T}\right)
\end{equation}
\begin{equation}
\left(\frac{\delta r_*}{r_*}\right)_{\Delta x} \simeq 0.42
\left(\frac{\delta\Delta x }{\Delta x}\right)
\end{equation}
\begin{equation}
\left(\frac{\delta r_*}{r_*}\right)_{B_{25}}=(-0.10+0.42 f)
\left(\frac{\delta B_{25}}{B_{25}}\right)
\end{equation}
\begin{equation}
\left(\frac{\delta r_*}{r_*}\right)_{a_{L,10}} = (-0.80 - 0.12 f)
\left(\frac{\delta a_{L,10} }{a_{L,10}}\right)
\end{equation}
\begin{equation}
\left(\frac{\delta r_*}{r_*}\right)_{S_{\nu,10}}= (-0.36 + 0.12 f)
\left(\frac{\delta S_{\nu,10} }{S_{\nu,10}}\right)
\end{equation}
\begin{equation}
\left(\frac{\delta r_*}{r_*}\right)_{b=1}=(-1.375 + 0.83 f)
\left(\frac{\delta b}{b}\right)
\end{equation}
\begin{equation}
\left(\frac{\delta r_*}{r_*}\right)_{b=0.25}=(-0.51 + 0.83 f)
\left(\frac{\delta b}{b}\right)
\end{equation}

The largest contribution to the uncertainty in $r_*$ is usually 
due to the uncertainty in the bridge width $a_{L,10}$.

\begin{deluxetable}{lcccccc}
\tablecaption{\label{tab:arch}VLA Archive Data Sets}
\tablehead{
\colhead{} & \colhead{} & \colhead{} & \colhead{Listed} &
\colhead{Frequency} & \colhead{Array} & \colhead{} \\
\colhead{Source} & \colhead{Band}  & \colhead{Program ID} & 
\colhead{Observer} 
& \colhead{(GHz)} & \colhead{Config.} & \colhead{Date}  
}
\startdata

3C 244.1 & L & POOL & Pooley, G. & 1.411 & B & 09/17/82 \\
 & C & AF213 & Fernini, I. & 4.872 & B & 12/23/91 \\
3C 337  & L & AR123 & Rudnick, L. & 1.452 & A & 02/21/85 \\
 & C & AR123 & Rudnick, L. & 4.885 & B & 06/01/85 \\
3C 325  & L & AV153 & van Breugel, W. & 1.465 & A & 12/05/88 \\
 & C & AF213 & Fernini, I. & 4.885 & B & 12/23/91 \\
3C 194  & L & AV164 & van Breugel, W. & 1.465 & A & 05/11/90 \\
 & C & AV164 & van Breugel, W. & 4.885 & A & 05/11/90 \\
3C 324  & L & AR123 & Rudnick, L. & 1.452 & A & 02/21/85 \\
 & C & AR123 & Rudnick, L. & 4.885 & B & 06/01/85 \\
3C 437  & L & AV164 & van Breugel, W. & 1.465 & A & 05/11/90 \\
 & C & AV164 & van Breugel, W. & 4.885 & B & 04/21/89 \\

\enddata
\end{deluxetable}

\begin{deluxetable}{llccccc}
\tablecaption{\label{tab:dstab}Radio Galaxies with $D_{\star}$ Presented Here}
\tablehead{
\colhead{}  & \colhead{}  & \colhead{}  &
\colhead{$\theta$}      &
\multicolumn{2} {c} {$r_{\star}$ \tablenotemark{a}}     &
\colhead{$D_{\star}$\tablenotemark{a}}  \\
\cline{5-6}
\colhead{Source}        &
\colhead{$z$}   &
\colhead{Bin}   &
\colhead{(arcsec)}      &
\colhead{($h^{-1}$ kpc)}        &
\colhead{($h^{-1}$ kpc)}        &
\colhead{($h^{-1}$ kpc)}        }
\startdata

 3C 244.1   & 0.43 & 2        & 50.8  & $151\pm16$    & $135\pm16$    & $286\pm22$           \\
 3C 337     & 0.63 & 3        & 43.5  & $150\pm24$    & $65\pm7$      & $214\pm25$           \\
 3C 325     & 0.86 & 3        & 15.8  & $164\pm52$    & $66\pm14$     & $230\pm54$           \\
 3C 194     & 1.19 & 4        & 14.2  & $105\pm22$    & $92\pm15$     & $197\pm27$           \\
 3C 324\tablenotemark{b}& 1.21 & 5        & 10.2  & $75\pm18$     & \nodata       & $149\pm35$           \\
 3C 437     & 1.48 & 5        & 36.7  & $55\pm6$      & $48\pm5$      & $103\pm8$            \\

\enddata
\tablenotetext{a}{Computed assuming $\beta=1.75$, $\Omega_0=0.1$,
$\Omega_{\Lambda}=0$, $b=0.25$ and not including the $\alpha-z$
correction.}
\tablenotetext{b}{$r_{\star}$ for only one bridge.}
\end{deluxetable}

\begin{deluxetable}{ccccccc}
\tablecaption{The Average Lobe-Lobe Sizes for 
Powerful 3CR Radio Galaxies. \label{tab:phys}}
\tablehead{
\colhead{} & \colhead{} & \colhead{} & \colhead{} & 
\multicolumn{3}{c}{$\langle D \rangle$ ($h^{-1}$ kpc)} \\  
\cline{5-7} \\
\colhead{Bin} & \colhead{$z$ Range} & \colhead{Sources} & \colhead{} &
\colhead{$\Omega_o=1.0$, $\Omega_{\Lambda}=0.0$} &
\colhead{$\Omega_o=0.1$, $\Omega_{\Lambda}=0.0$} &
\colhead{$\Omega_o=0.1$, $\Omega_{\Lambda}=0.9$} 
}
\startdata

1 & 0.0-0.3 & 3  & & 66$\pm$14  & 68$\pm$14  & 72$\pm$13  \\
2 & 0.3-0.6 & 13 & & 202$\pm$45 & 224$\pm$50 & 259$\pm$57 \\
3 & 0.6-0.9 & 23 & & 148$\pm$17 & 174$\pm$20 & 209$\pm$24 \\
4 & 0.9-1.2 & 16 & & 107$\pm$24 & 133$\pm$29 & 165$\pm$36 \\
5 & 1.2-1.6 & 9  & & 91$\pm$32  & 122$\pm$43 & 152$\pm$53 \\
6 & 1.6-2.0 & 6  & & 66$\pm$19  & 92$\pm$26  & 114$\pm$33 

\enddata
\end{deluxetable}

\begin{deluxetable}{lllcccccc}
\tablecaption{\label{tab:natab}Summary of New Source Properties}
\tablehead{
\colhead{Source} &
\colhead{$z$}   &  \colhead{$r$\tablenotemark{a}} &
\colhead{$a_L$\tablenotemark{b}}        &
\colhead{$B_{10}$\tablenotemark{c}}     &
\colhead{$B_{25}$\tablenotemark{d}}     &
\colhead{$v_L$\tablenotemark{e}}        &
\colhead{$n_a$\tablenotemark{f}}        &
\colhead{$\log L_j$\tablenotemark{g}}
}

\startdata
244.1  & 0.43  & 101  & $5.5\pm0.3$   
& $3.5\pm0.2$  & $2.5\pm0.2$   & $1.1\pm0.2$
 & $1.0\pm0.4$   & $44.35\pm0.10$         \\
           &       & 91.6 & $4.7\pm0.4$   
& $5.3\pm0.3$  & $3.2\pm0.2$   & $1.7\pm0.3$
 & $1.0\pm0.4$   & $44.75\pm0.11$         \\
337    & 0.63  & 125 & $6.2\pm0.5$   
& $3.5\pm0.2$  & $2.3\pm0.1$   & $1.5\pm0.5$
 & $0.6\pm0.4$   & $44.57\pm0.15$         \\
           &       & 69.8 & $10.5\pm0.4$  
& $3.8\pm0.2$  & $2.8\pm0.2$   & $1.1\pm0.2$
 & $1.2\pm0.4$   & $44.99\pm0.10$         \\
 325    & 0.86  & 49.0 & $3.3\pm1.1$   
& $6.8\pm0.9$  & $4.0\pm0.3$   & $2.0\pm0.5$
 & $1.2\pm0.6$   & $44.73\pm0.23$         \\
           &     & 34.7  & $4.4\pm0.8$   
& $12.7\pm1.0$ & $4.6\pm0.3$   & $3.0\pm0.8$
 & $1.9\pm1.0$   & $45.70\pm0.17$         \\
 194    & 1.19 & 31.6  & $4.5\pm1.0$   
& $7.1\pm0.6$  & $5.0\pm0.3$   & $1.8\pm0.3$
 & $1.6\pm0.6$   & $45.00\pm0.16$         \\
           &      & 50.0 & $5.6\pm0.8$   
& $6.7\pm0.5$  & $5.6\pm0.4$   & $2.0\pm0.3$
 & $1.2\pm0.4$   & $45.16\pm0.13$         \\
 324    & 1.21  & 22.1 & $4.5\pm0.9$   
& $10.3\pm0.9$ & \nodata       & $2.1\pm0.7$
 & $2.4\pm1.5$   & $45.38\pm0.19$         \\
 437    & 1.48  &  106 & $11.4\pm0.8$  
& $6.9\pm0.4$  & $4.2\pm0.2$   & $4.6\pm0.8$
 & $0.2\pm0.1$   & $46.18\pm0.10$         \\
           &     & 101  & $10.5\pm0.8$  
& $9.4\pm0.6$  & $6.7\pm0.4$   & $6.0\pm1.0$
 & $0.2\pm0.1$   & $46.49\pm0.10$         \\

\enddata
\tablenotetext{a}{Core-hotspot distance in $h^{-1}$ kpc. }
\tablenotetext{b}{Lobe radius, 10 $h^{-1}$ kpc behind
hot spot, in $h^{-1}$ kpc. }
\tablenotetext{c}{Minimum energy magnetic field, 10 $h^{-1}$
kpc behind hot spot, in $h^{2/7} 10^{-5}$G.}
\tablenotetext{d}{Minimum energy magnetic field, 25 $h^{-1}$
kpc behind hot spot, in $h^{2/7} 10^{-5}$G.}
\tablenotetext{e}{Lobe advance speed, in $10^{-2}c$.}
\tablenotetext{f}{Ambient gas density, in $10^{-3}
h^{n} \mbox{cm}^{-3}$, where $n = 12/7$ for $B >> B_{MB}$ and 
$n= 20/7$ for $B << B_{MB}$.}
\tablenotetext{g}{Logarithm of the luminosity in directed kinetic
energy, in $h^{-2} \mbox{erg s}^{-1}$.}
\end{deluxetable}

\clearpage

\clearpage

\clearpage

\clearpage

\clearpage

\clearpage

\clearpage

\clearpage

\clearpage

\clearpage

\clearpage

\clearpage

\clearpage

\end{document}